\documentclass[a4paper,11pt]{article}
\pdfoutput=1
\usepackage{tablefootnote}
\usepackage{scrextend}

\usepackage{jheppub}
\pdfoutput=1
\usepackage{epsf}
\usepackage{epsfig}
\usepackage{subfig}
\usepackage{mathtools}    
\usepackage{float}
\usepackage{multirow}
\usepackage{ mathrsfs }
\usepackage{nicefrac}
\usepackage{epstopdf}
\usepackage{slashed}
\usepackage[dvipsnames]{xcolor}
\usepackage{url}
\usepackage{breqn}
\usepackage{amsmath}   
\usepackage[export]{adjustbox} 
\usepackage{chemformula}
\usepackage{multicol}
\usepackage{placeins}
\usepackage{braket}
\usepackage{subdepth}
\usepackage{color}
\usepackage{booktabs}
\usepackage{siunitx} 

\newcommand{\ifb}{\text{ fb}^{-1}}
\newcommand{\tev}{\text{TeV}}

\usepackage{soul}

\graphicspath{ {./Figures/} }

\title{{Renormalisation group evolution effects on global SMEFT analyses}}

\author{Riccardo Bartocci, Anke Biekötter, Tobias Hurth} 

\affiliation{PRISMA+ Cluster of Excellence \& Institute of  Physics (THEP) \& Mainz Institute for Theoretical Physics, Johannes Gutenberg University, D-55099 Mainz, Germany}

\abstract{
Global analyses in the Standard Model Effective Field Theory (SMEFT) framework serve as a tool to probe potential directions of new physics. To break degeneracies between the Wilson coefficients of the SMEFT, it is essential to combine observables from various experiments. 
Since different observables entering global fits may be measured at different energy scales, it becomes increasingly important to account for this fact through the renormalisation group evolution (RGE) of the Wilson coefficients. 
In this work, we investigate the effects of the RGE on a global SMEFT fit under the assumption of a $U(3)^5$ symmetry within the minimal flavour violation framework. 
We comment on the role of next-to-leading order SMEFT predictions for breaking potential degeneracies between Wilson coefficients arising as a result of RGE effects. 
}

\preprint{
\begin{minipage}{3cm}
\small
\flushright
MITP-24-087
\end{minipage}}

\begin{document}

\maketitle
%%%%%%%%%%%%%%%%%%%%%%%%%%%%%%%%%%%%%%%%%%%%%%%%%%%%%%%%%%%%%%%%%%%%%%%%%%%%%%%%%%%%%%%%%%%%%%%%%%%%%%%%%%%%%%%%%%%%%%%%%%%%%%%%%%%%%%%%%%%%%%%%%%%%%%%%%%%%%%%%%%%%%%
%%%%%%%%%%%%%%%%%%%%%%%%%%%%%%%%%%%%%%%%%%%%%%%%%%%%%%%%%%%%%%%%%%%%%%%%%%%%%%%%%%%%%%%%%%%%%%%%%%%%%%%%%%%%%%%%%%%%%%%%%%%%%%%%%%%%%%%%%%%%%%%%%%%%%%%%%%%%%%%%%%%%%%

%----------------------------------------------
\newpage
\section{Introduction}
\label{sec:introduction}
%----------------------------------------------

In the absence of direct new physics (NP) discoveries, Standard Model Effective Field Theory (SMEFT)~\cite{Buchmuller:1985jz,Wilczek:1977pj,Grzadkowski:2010es,Brivio:2017vri} provides a powerful framework to describe deviations from the Standard Model (SM) expectation. 
Its main advantages lie in the minimal assumptions on the realisation of NP at a high and not currently experimentally accessible scale $\Lambda$ and its systematic expansion in inverse powers of $\Lambda$ and couplings. 

Global SMEFT analyses play a crucial role in collecting experimental constraints on potential directions of new physics and assessing their status. They have been performed for various sectors of experimental observables including low-energy~\cite{Falkowski:2017pss,Falkowski:2019xoe,Falkowski:2023klj}, flavour~\cite{Aoude:2020dwv,Bruggisser:2021duo,Bruggisser:2022rhb,Grunwald:2023nli}, electroweak~\cite{Biekoetter:2018ypq,Kraml:2019sis,Dawson:2020oco,Almeida:2021asy,Anisha:2021hgc} and top~\cite{Buckley:2015lku,Aguilar-Saavedra:2018ksv,Brivio:2019ius,Bissmann:2019gfc,Durieux:2019rbz}  data as well as combinations thereof~\cite{Ellis:2020unq,Ethier:2021bye,Garosi:2023yxg,Bartocci:2023nvp,Celada:2024mcf}. 
To probe all directions of the NP parameter space in the SMEFT framework and break potential degeneracies between coefficients, it is crucial to combine observables from various experiments. For instance, electroweak precision observables (EWPO), Higgs and diboson data constrain different combinations of the Wilson coefficients contributing to the modification of $Zq\bar{q}(h)$ couplings. As another example, degeneracies between semileptonic operators in Drell-Yan~(DY) are broken in atomic parity violation and parity violation electron scattering~\cite{Boughezal:2021kla}.

Current global fits of LHC data are typically performed at the scale of the observables and assume that renormalisation group evolution (RGE) effects~\cite{Jenkins:2013zja,Jenkins:2013wua,Alonso:2013hga} between the various observables can be neglected. The inclusion of RGE effects between the scale of NP $\Lambda$ and the scale of the observables, as needed for the interpretation in terms of NP models, is not generally given.
However, the combination of observables measured at different scales within one global fit creates the need for the inclusion of RGE effects within the fit. 
While this is already standard in the analysis of flavour data, the relevance of RGE effects for {high energy} LHC and future collider SMEFT analyses has only recently received more attention~\cite{Battaglia:2021nys, Aoude:2022aro,DiNoi:2023onw,DiNoi:2024ajj,Heinrich:2024rtg,Maltoni:2024dpn,Asteriadis:2024qim}, finding a sizeable impact of the running and mixing of the Wilson coefficients. Several studies have also highlighted the constraining power of present and future EWPO on four-quark operators and in particular those involving top quarks~\cite{Greljo:2023bdy,Allwicher:2023shc,Haisch:2024wnw,Gargalionis:2024jaw}. 

The aim of this paper is to assess the relevance of RGE effects on global SMEFT analyses including observables measured at a wide range of energy scales, from the sub-GeV level to kinematic distributions at several TeV. 
We perform an RGE improved global analysis of all CP even Wilson coefficients of the Warsaw basis~\cite{Grzadkowski:2010es} obeying a $U(3)^5$ SMEFT symmetry at the high scale {at tree level.
This flavour assumption within the Minimal Flavour Violation (MFV)  framework~\cite{Gerard:1982mm,Chivukula:1987py,Hall:1990ac,DAmbrosio:2002vsn}  reduces the number of independent Wilson coefficients to 41. It  is motivated by the fact that flavour observables push the appearance of flavour-violating operators far above the TeV scale~\cite{Calibbi:2017uvl,Silvestrini:2018dos}. 
{The requirement of a $U(3)^5$ symmetry for the SMEFT operators (at tree level) corresponds to a rather specific class of new physics models and reflects the minimal and unavoidable amount of  flavour changing neutral currents (FCNCs) at the electroweak/low scale.
Thereby, it can serve as a test case for a global analysis of an operator set motivated by a symmetry assumption with a manageable number of free coefficients. }
The lessons learned in this work are not expected to result in significantly different findings than a less restrictive flavour assumptions at the high scale~\cite{Faroughy:2020ina,Greljo:2022cah}.

We analyse the additional sensitivity gained on the Wilson coefficients provided by RGE effects on a single-parameter fit level, finding that EWPO, parity violation experiments~(PVE) and DY provide additional constraints on four-quark operators. 
Moreover, we investigate RGE effects on a global SMEFT fit and discuss the origins of modifications to the limits of individual Wilson coefficients, focusing in particular on the cross talk of different observables and the new degeneracies appearing. 
We already anticipate that the largest effects are observed for the Wilson coefficients of four-quark operators and those that have strong correlations with them, in particular operators contributing to modified $Zq\bar{q}(h)$ couplings. 

The paper is organised as follows. In Section~\ref{sec:basics}, we introduce the SMEFT framework and notations and give details on the observables and corresponding SMEFT predictions included in our fit. 
In Section~\ref{sec:inclusion_RGE}, we give details on our treatment of RGE effects on the SMEFT predictions. 
In Sections~\ref{sec:single_parameter} and~\ref{sec:global}, we present limits on the Wilson coefficients after the inclusion of RGE effects from a single-parameter fit and global fit, respectively. 
We conclude in Section~\ref{sec:conclusions}. 
In Appendix~\ref{app:PCA}, we present a principal component analysis of our global fit. 
Appendix~\ref{app:diffs_prev} is dedicated to a discussion of difference with respect to our previous work~\cite{Bartocci:2023nvp}. In Appendices~\ref{app:operators} and~\ref{app:observables}, we list the operators and observables included in our analysis. 
We present numerical results for the limits and correlation matrices in Appendix~\ref{app:num_res}.

%----------------------------------------------
\section{SMEFT assumptions and fit inputs}
\label{sec:basics}
%----------------------------------------------
The SMEFT Lagrangian, truncated at dimension six, is given by 
\begin{equation}
\mathcal{L}_{\text{SMEFT}}=\mathcal{L}_{\text{SM}}+\sum_i \frac{C_i}{\Lambda^2} \, Q_i,
\label{eq:SMEFT}
\end{equation}
where $C_i$ denote the Wilson coefficients of the operators $Q_i$ in the Warsaw basis~\cite{Grzadkowski:2010es} and $\Lambda$ denotes the NP scale, 
which we set to $\Lambda=4\,\tev$ throughout our work. 
We truncate all SMEFT predictions at linear order in the Wilson coefficients, neglecting quadratic contributions which are suppressed by $\Lambda^{-4}$ and therefore formally of the same order as dimension-eight contributions. We employ the electromagnetic coupling constant, the mass of the $Z$ boson and the Fermi constant $\{\alpha,\,M_Z,\,G_F\}$ as our electroweak input parameters.

In order to reflect the minimal amount of FCNCs and CP violation observed at the electroweak scale, we make assumptions on the flavour structure of the NP interactions and their CP nature.
We consider a $U(3)^5$ symmetry of the SMEFT as initial condition at the high scale at tree level, specifically 
\begin{equation}
U(3)^5=U(3)_\ell \times U(3)_q\times U(3)_e\times U(3)_u\times U(3)_d,
\end{equation}
where $\{\ell, q, e, u, d\}$ represent the SM fermions~\cite{Faroughy:2020ina}. 
This corresponds to the assumption that all new interactions of a model beyond the SM match onto flavour symmetric dimension-six operators at tree level.
In addition, we restrict ourselves to CP-even interactions, as CP-odd ones can be much better constrained using dedicated search strategies~\cite{Ferreira:2016jea,Brehmer:2017lrt,Bernlochner:2018opw,Englert:2019xhk,Cirigliano:2019vfc,Biekotter:2020flu,Biekotter:2021int,Bakshi:2021ofj,Degrande:2021zpv,Bhardwaj:2021ujv,Hall:2022bme,Thomas:2024dwd}.
These two assumptions reduce the list of independent Wilson coefficients at dimension six to 41. We list the corresponding operators in Tab.~\ref{tab:basis}. 
As already pointed out in the introduction, FCNCs at the electroweak scale are still predicted within our SMEFT flavour assumption, as a result of loop matching and RGE effects.
However, all additional Wilson coefficients induced by these effects explicitly depend on the 41 parameters considered in our analysis.

%----------------------------------------------
\subsection{Datasets}
\label{subsec:datasets}
Our analysis combines observables measured at a wide range of energy scales, from the sub-GeV level to kinematic distributions at several TeV.
Specifically, we include data from EWPO, diboson, Higgs, top, low-energy parity violation experiments (PVE), lepton scattering, flavour, DY as well as dijet+photon production.
As this dataset almost exactly agrees with the one used for our previous study~\cite{Bartocci:2023nvp}, we refer to this study for a detailed description of the included datasets and the Wilson coefficients constrained in each set 
and only comment on the differences with respect to~\cite{Bartocci:2023nvp} in this section.
We also list the observables included in our fits in Appendix~\ref{app:observables}, 
in Tables~\ref{tab:obset}, \ref{tab:obset_top}, \ref{tab:obset_top2_DY_dijet}, and \ref{tab:obset_PVE_flavour}.

In addition to our previous dataset~\cite{Bartocci:2023nvp}, we incorporate observables from $\beta$-decay and semileptonic meson decays as a single pseudo-observable, $\Delta_{\text{CKM}}$, which represents the unitarity of the CKM matrix~\cite{Gonzalez-Alonso:2016etj,Falkowski:2020pma,Cirigliano:2022qdm,ThomasArun:2023wbd,Cirigliano:2023nol}.
We showed that this pseudo-observable mainly exhibits sensitivity to $C_{lq}^{(3)}$ in a global fit~\cite{Bartocci:2023nvp}. 
Furthermore, we also include kinematic distributions of LHC diboson measurements from ATLAS in the $WZ$~\cite{ATLAS:2019bsc} and $WW$~\cite{ATLAS:2019rob} final states. 
The impact of these additional datasets on the fit is discussed in Appendix~\ref{app:diffs_prev}.

%-------------------------------------------------
%-------------------------------------------------
\subsection{SMEFT predictions}
\label{sec:SMEFT_pred}
%-------------------------------------------------
We utilise SMEFT predictions from Ref.~\cite{Biekotter:2023xle} (EWPO), \texttt{fitmaker}~\cite{Ellis:2020unq} (Higgs, top), Ref.~\cite{Anisha:2021hgc} (Higgs), \texttt{SMEFiT}~\cite{Ethier:2021bye} ($t\bar{t}$ at NLO), \texttt{flavio}~\cite{Straub:2018kue} (flavour), Ref.~\cite{Falkowski:2017pss, Falkowski:2015krw} (PVE, lepton scattering), \texttt{high-pT}~\cite{Allwicher:2022mcg} (Drell-Yan), Ref.~\cite{Bartocci:2023nvp} (dijet+photon production) and Ref.~\cite{Cirigliano:2023nol} ($\Delta_{\text{CKM}}$).
Some of these predictions assume flavour universality for the Wilson coefficients. To correctly account for the flavour dependence of the RGE effects, we reintroduce flavour coefficients in the predictions as follows. 
We assume that quark couplings to the Higgs boson as well as dipole operators are dominated by the third-generation quarks 
\begin{align}
    C_{uG} \to C_{\substack{uG\\33}} , \quad 
    C_{uW} \to C_{\substack{uW\\33}} , \quad
    C_{uB} \to C_{\substack{uB\\33}} , \quad
    C_{uH} \to C_{\substack{uH\\33}} , \quad 
    C_{dH} \to C_{\substack{dH\\33}} \, , 
\end{align}
where the notation $C_{\substack{x\\ij}}$ is used for the flavour indices $i$ and $j$.
For top-sector predictions from \texttt{fitmaker}, third-generation indices are given explicitly and we assume that first- and second generation quarks equally contribute to the light-quark  and lepton contributions, e.g.\ $C_{Hl}^{(3)} \to 1/2 (C_{\substack{Hl\\11}}^{(3)} + C_{\substack{Hl\\22}}^{(3)})$.

For dijets+$\gamma$ production, we regenerate the predictions including flavour indices using \texttt{SMEFTsim}~\cite{Brivio:2020onw}\footnote{We attach these as an ancillary file to the arXiv submission.}. We give more details on this in Appendix~\ref{app:dijet_plus_photon}. 
The EWPO, DY, lepton scattering, PVE and flavour predictions are already flavour general, so no adjustments are needed. 

With respect to~\cite{Bartocci:2023nvp}, we have also fixed a bug in our translation of the Wilson coefficients from the basis of \texttt{SMEFTatNLO}~\cite{Degrande:2020evl} to the Warsaw basis.\footnote{We had not taken into account differences between the independent basis formulation of the Warsaw basis and its symmetric counterpart. Moreover, we copied an inconsistency in the relation between Wilson coefficients in the basis used by \texttt{SMEFTatNLO} and the Warsaw basis from~\cite{Ellis:2020unq}.} 
This mistake changed the normalisation of the Wilson coefficients $C_{qu}^{(x)}, \, C_{qd}^{(x)}$ and $C_{ud}^{(x)}$ with $x=1,\,8$ in the top-sector SMEFT predictions by a factor of two and hence affected the top-dominated bounds by this factor. The discussion of the interplay of different datasets remains unchanged.
We have also adjusted the sign of $C_G$ such that all the Wilson coefficient definitions now agree with the ones employed in \texttt{SMEFTsim}.

%--------------------------------------
\section{RGE effects}
\label{sec:inclusion_RGE}
The energy scale of the observables included in our fit ranges from below 1~GeV, as in the case of PVE, over the mass of the $Z$ boson for EWPO, up to 2.3~TeV for kinematic distributions in DY and \(t\bar{t}\) production.
To account for the scale dependence of the Wilson coefficients and facilitate the reinterpretation in terms of concrete models, we incorporate RGE effects into our SMEFT predictions. 
The importance of these effects for SMEFT analyses has been highlighted in~\cite{Battaglia:2021nys, Aoude:2022aro,DiNoi:2023onw,DiNoi:2024ajj,Heinrich:2024rtg}.

In order to account for RGE effects, we need to assign a scale to each of our observables. 
As a baseline, we define the scale as the mass of the $Z$ boson, the mass of the Higgs boson $m_h$ and the mass of the top quark $m_t$ in EWPO, Higgs and top-quark observables, respectively.   
In energy-dependent kinematic distributions, we use the invariant mass of the final state particles or the squared sum of a particle mass and its transverse momentum, $\sqrt{m_i^2 + p_{T,i}^2}$, as a scale measure. 
For bins in distributions, we use the lower edge of each bin as a measure of the scale of the process, as most distributions fall off quickly with enhanced energy. 
The scale assigned to each observable is explicitly listed in the last column of Tables~\ref{tab:obset}, \ref{tab:obset_top}, \ref{tab:obset_top2_DY_dijet}, and \ref{tab:obset_PVE_flavour} of Appendix~\ref{app:observables}.

In the EFT framework, accounting for RGE effects is essential for the consistent perturbative treatment of quantum corrections at the low scale.
To perform the RG evolution, we employ the \texttt{DsixTools} package~\cite{Celis:2017hod,Fuentes-Martin:2020zaz,Jenkins:2017jig,Jenkins:2017dyc,Dekens:2019ept}. 
Resummed leading-log RGE effects are included via its evolution matrix approach which allows us to express the evolution of the Wilson coefficients from a high scale $\Lambda$ to a lower scale $\mu$ as
\begin{equation}
    C_i (\mu) = U_{ij}\left(\mu, \Lambda \right) \,  C_j (\Lambda) \, 
\end{equation}
where $U_{ij}$ denotes the evolution matrix. 
As already pointed out, we choose a fixed value of $ \Lambda = 4$~TeV throughout our study. This value is approximately twice the scale of the highest-energy observable included in our fit to ensure the validity of the EFT approach.
Note that at scales $\mu < \Lambda$, the set of Wilson coefficients included in our analysis is no longer constrained to those obeying a $U(3)^5$ symmetry as flavour non-conserving interactions are still allowed within the SM. 
However, all of the Wilson coefficients induced through the RGE  are dependent on the 41 coefficients defined at the NP scale. 
Therefore, they do not correspond to independent NP directions in our fit and cannot induce any flat directions.

Taking into account RGE effects represents a crucial step forward in enhancing the precision of global analyses. In addition, the inclusion of RGE contributions serves as a probe to assess the importance of incorporating NLO effects in global fits. 
Although significant progress is being made on the automatization of the calculation of NLO predictions, 
automated results are currently only available for NLO QCD corrections~\cite{Degrande:2020evl}. 
Therefore, incorporating RGE effects in global SMEFT fits allows probing the relevance of NLO corrections consistently.

%-------------------------------------------------
\section{Single-parameter limits from RGE effects}
\label{sec:single_parameter}
%-------------------------------------------------
\begin{figure}[t]
    \centering
\includegraphics[scale=0.5,valign=t]{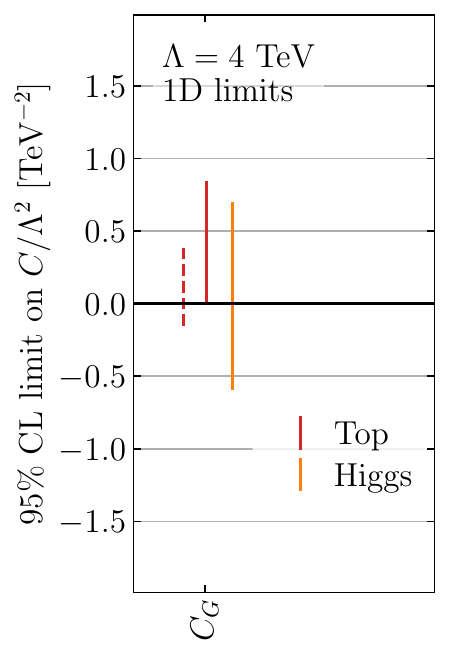}
\includegraphics[scale=0.5,valign=t]{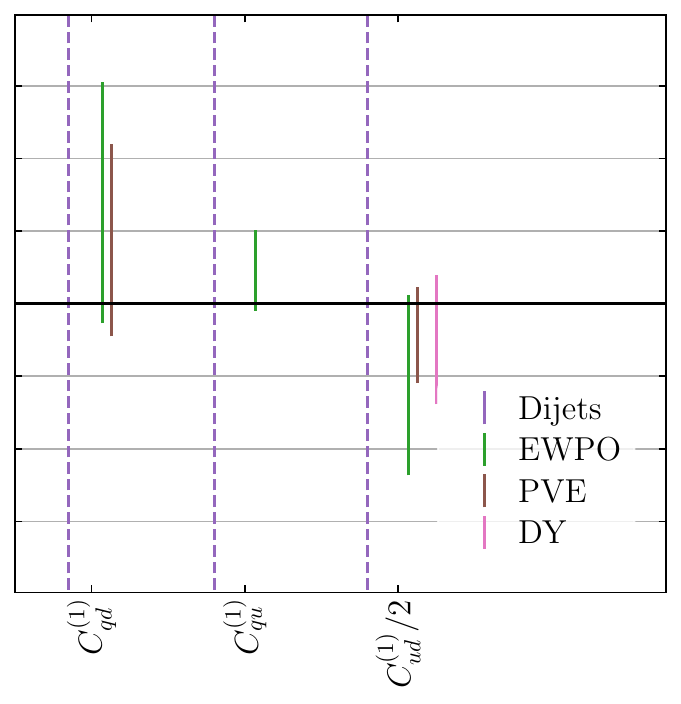}\\
\includegraphics[height=6cm]{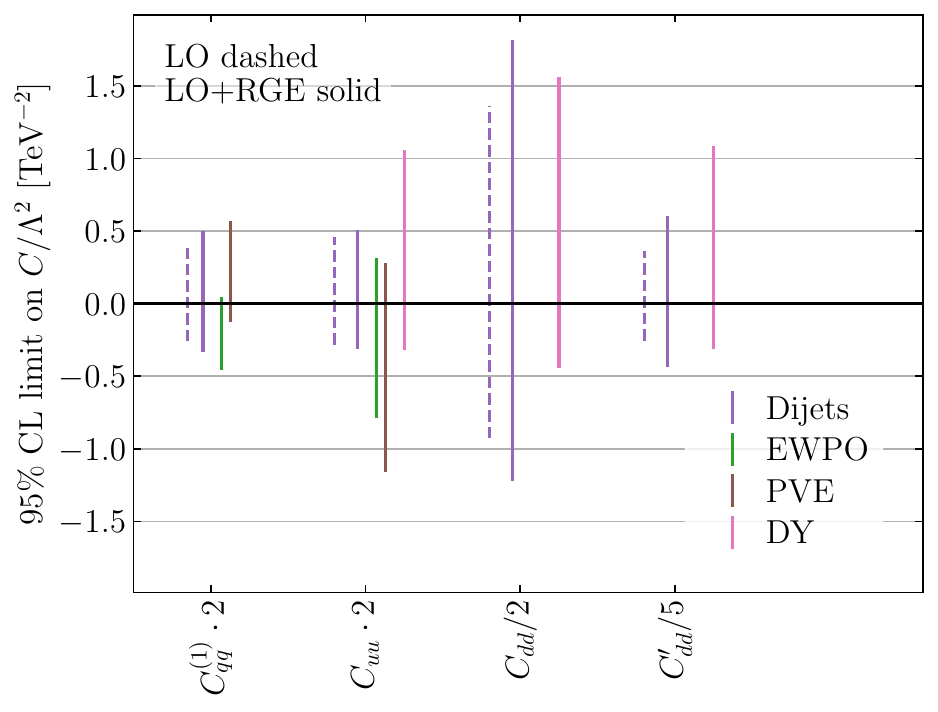}\\
    \caption{Single-parameter limits on the Wilson coefficients before (dashed) and after (solid) the inclusion of RGE effects from different datasets. 
    We only display the strongest bound for the LO case and the LO+RGE bounds which are at most a factor two wider than the strongest bound in this setting. 
    The different panels separate operators involving only gauge fields, four-quark operators with a suppressed interference in dijets+$\gamma$ production and additional four-quark operators. }
    \label{fig:1D_limits}
\end{figure}
The inclusion of RGE mixing effects implies that operators defined at the high scale mix into operators defined at lower scales. 
Consequently, datasets that at LO do not constrain certain coefficients can provide significant constraints through RGE effects. 
These newly introduced constraints can be competitive or even stronger than those present before RGE effects are included, in cases where LO constraints are weak.

We present in Fig.~\ref{fig:1D_limits} the limits from single-parameter fits before (dashed lines) and after (solid lines) the inclusion of RGE effects for individual datasets. We show only the strongest bound for the LO case and the LO+RGE bounds that are at most a factor of two weaker than the strongest bound in this setting. For the coefficient $C_G$, the best constraints at LO and LO+RGE come from top quark processes. However, via mixing with the coefficient $C_{HG}$, the inclusion of RGE effects provides substantial additional constraints from Higgs data.

At LO, the only limits on the operators $C_{qd}^{(1)}$, $C_{qu}^{(1)}$, and $C_{ud}^{(1)}$ result from dijet$+\gamma$ data. As the dijet production diagrams including the corresponding SMEFT operators and the dominant SM diagram do not interfere, the Wilson coefficients are poorly constrained at LO.
When RGE effects are included, we observe additional limits from EWPO, PVE and DY on these coefficients.
The mixing of $C_{qd}^{(1)}$ and $C_{ud}^{(1)}$ with the semileptonic operator $C_{ed}$, is responsible for the bounds from PVE and DY. Despite the fact that these mixing effects are suppressed by the electromagnetic coupling constant $\alpha$, they still lead to competitive constraints at the single-parameter fit level. This is a reflection of the fact that currently no dataset probes these coefficients well at LO. 
The constraints from EWPO result from the mixing of the considered four-quark operators with $C_{Hu}$ and $C_{Hd}$, both of which are tightly constrained by EWPO.

The operator $C_{qq}^{(1)}$ is dominantly constrained by dijet data at LO. This constraint remains relevant even when RGE effects are included, but after RGE, the most stringent bound comes from EWPO due to significant mixing with $C_{Hq}^{(1)}$. 
An additional constraint arises from PVE due to the mixing of $C_{qq}^{(1)}$ and $C_{qe}$. 
The coefficient $C_{uu}$ mixes with operator $C_{Hu}$ and the semileptonic operators $C_{eu}$ and $C_{lu}$, leading to additional bounds from EWPO, DY and PVE. 
Nevertheless, its dominant constraint still arises from dijet data.
Finally, the operators $C_{dd}$ and $C_{dd}^\prime$, which at LO are constrained solely by dijet data, receive significant additional constraints from DY when mixing effects are included, even though the mixing of $C_{dd}$ and $C_{dd}^\prime$ with semileptonic operators is suppressed by $\alpha$. 
For $C_{dd}$, DY even provides the strongest single-parameter constraint.

%------------------------------------------
\section{Global analysis including RGE effects}
\label{sec:global}
%------------------------------------------
In this section, we present a global analysis based on RGE improved LO predictions. 
Following the procedure outlined in our previous analysis~\cite{Bartocci:2023nvp}, we derive limits on a specific Wilson coefficient while profiling over the remaining parameters using the toy Monte Carlo method. We include correlations between different observables, where known.

\begin{figure}
    \centering
    \includegraphics[height=6cm]{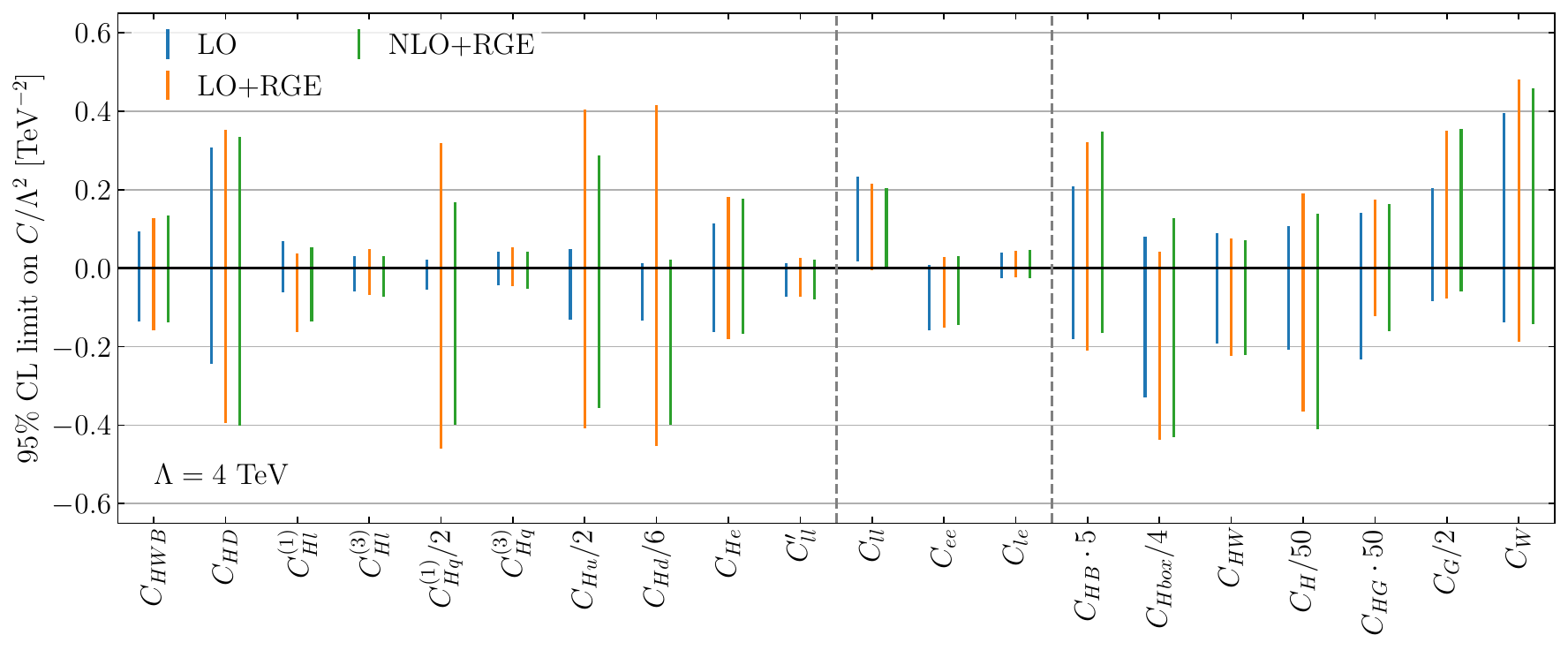}
    \includegraphics[height=6cm]{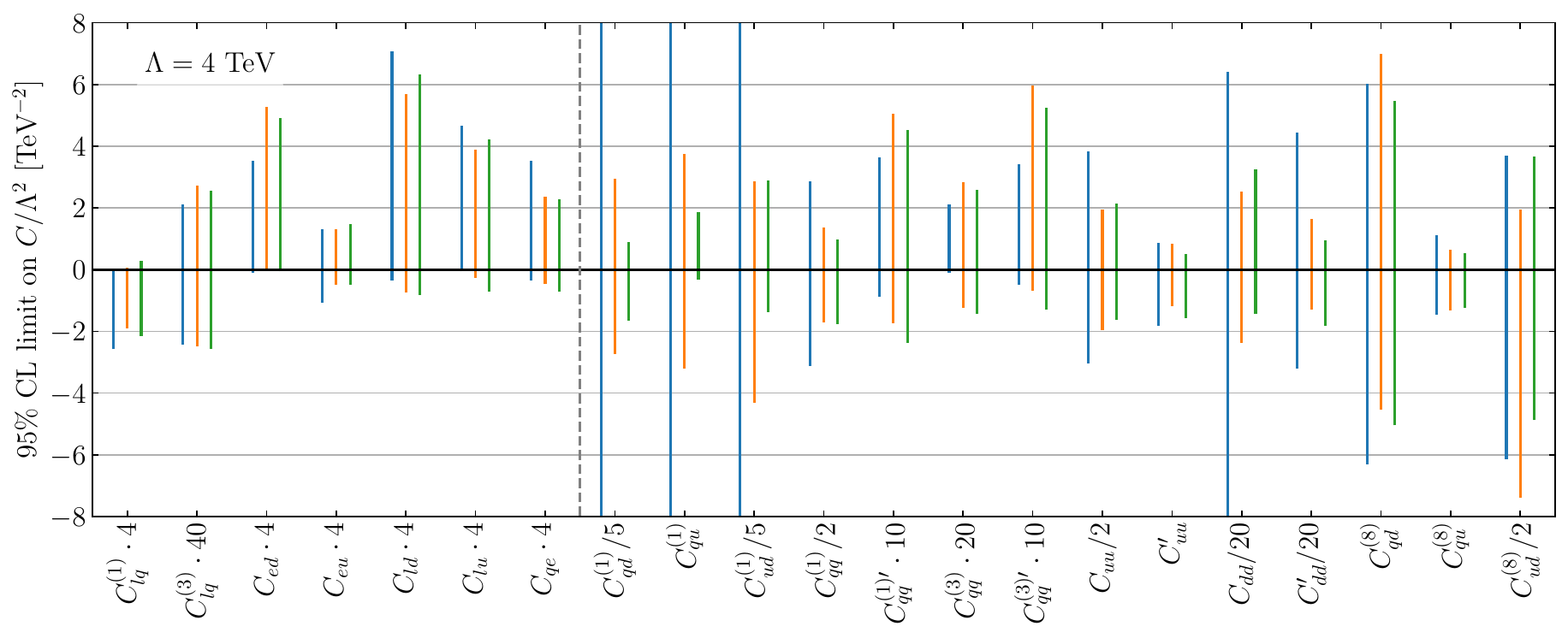}
    \caption{Comparison of the limits on the Wilson coefficients from a global fit purely based on LO predictions~(LO), with those including RGE effects (LO+RGE) and including partial NLO predictions (NLO+RGE). Note that some Wilson coefficients have been scaled by appropriate factors on the $x$ axis and that the $y$ axis is different between the two panels. }
    \label{fig:LO_NLO_RGE}
\end{figure}
In Figure~\ref{fig:LO_NLO_RGE}, we show the $95\%$ CL limits from a global fit on all $41$ Wilson coefficients.
We compare the bounds from the LO analysis (blue) and the RGE improved LO analysis (orange). 
We also show the limits from an RGE improved analysis based on partial NLO predictions (green), which we  discuss in Section~\ref{sec:NLO_effects}. 
Overall, we find that the inclusion of RGE effects helps remove the (almost) flat directions present in the LO fit. Specifically, it significantly improves the bounds on $C_{qd}^{(1)}$, $C_{qu}^{(1)}$, $C_{ud}^{(1)}$ and $C_{dd}$, which are only weakly constrained at LO.  To quantify this statement, we present the results of a principal component analysis (PCA) in Appendix~\ref{app:PCA}.
In the following, we discuss the main differences between the LO fit and RGE improved LO fit for different Wilson coefficient categories.

%----------------------------------------------
\paragraph{Operators contributing to EWPO at LO}
%---------
The global limits on $C_{Hq}^{(1)}$, $C_{Hu}$ and $C_{Hd}$, which contribute to modified $Zq\bar{q}(h)$ couplings, increase by factors $10.1$, $4.5$ and $5.9$, respectively. 
This is the result of significant correlations between the quark-gauge operators and four-quark operators induced by the RG evolution.
For $C_{Hq}^{(3)}$, which also contributes to modified $Wq\bar{q}'(h)$ interactions, correlations with four-quark operators are milder as a result of the additional constraints from $W$ decay and $Wh$ production.
\begin{figure}
    \centering
    \includegraphics[width=0.48\linewidth,valign=t]{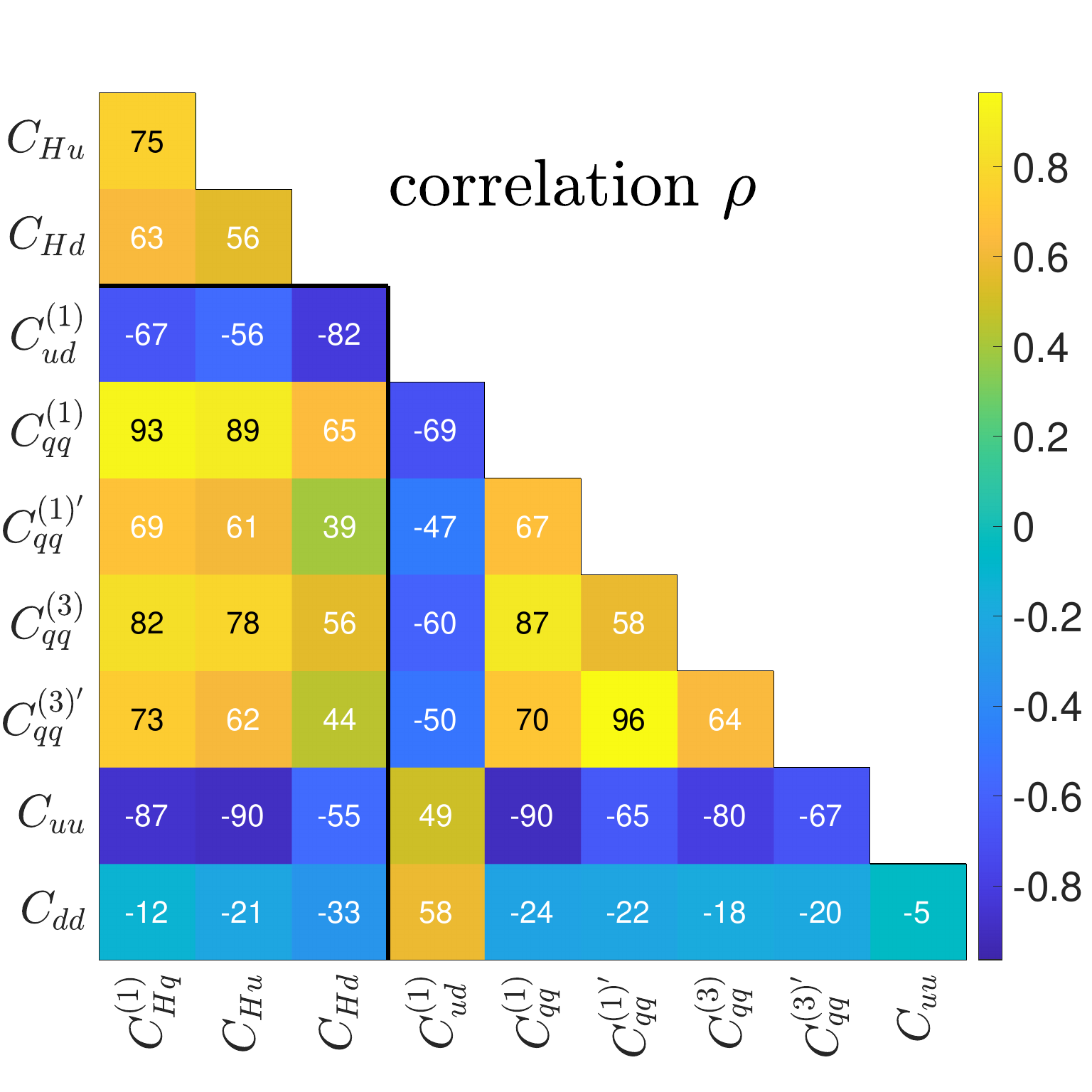}
    \quad
    \includegraphics[width=0.48\linewidth,valign=t]{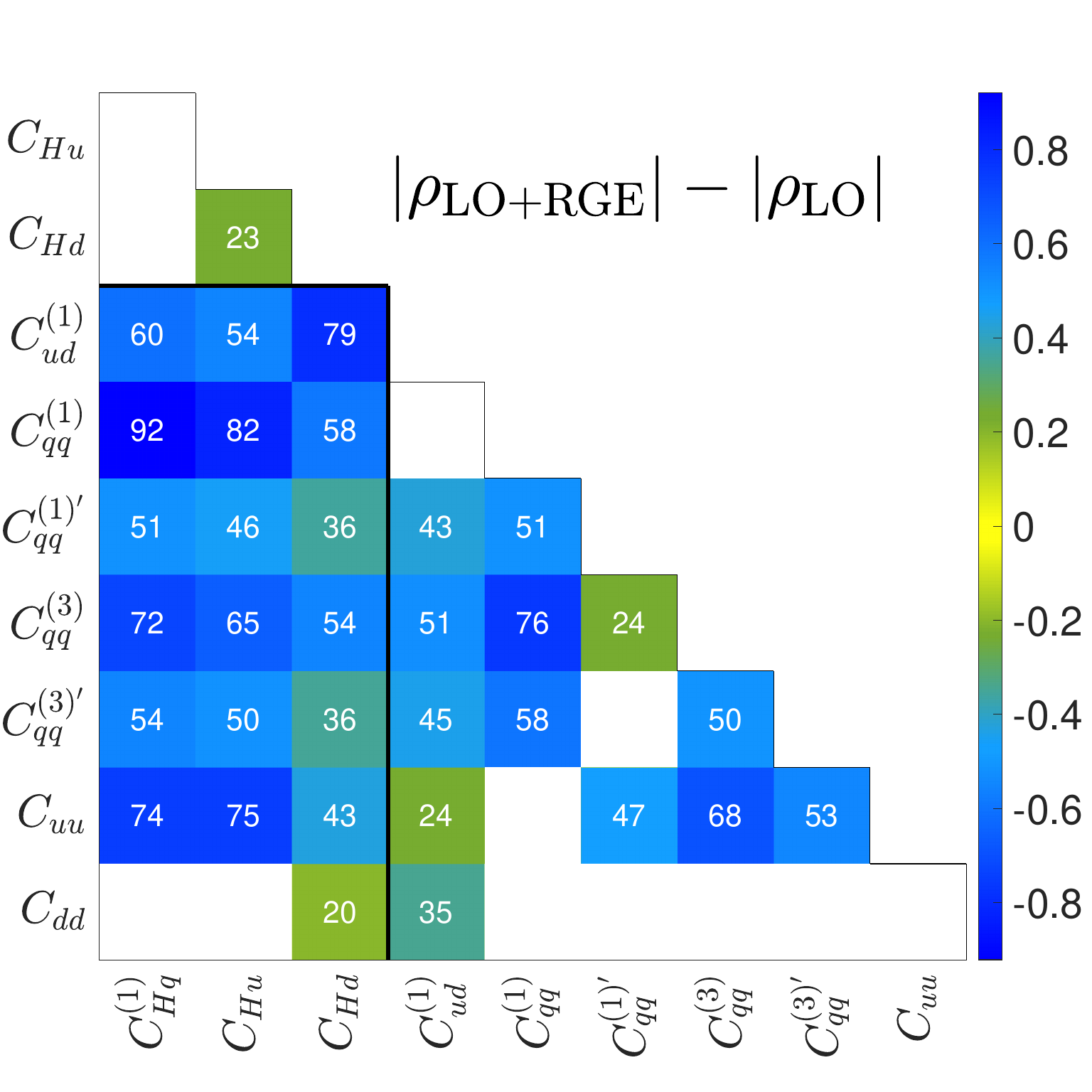}
    \caption{
    Left: Correlation $\rho$ (in percent) between selected Wilson coefficients in the LO+RGE fit. 
    Right: Difference between the absolute values of the correlations before and after the inclusion of RG evolution effects in the fit. Correlation changes below $20\%$ have been 
    omitted for better visibility of the dominant effects.
    }
    \label{fig:corr_diff_RG}
\end{figure}
In the top left panel of Fig.~\ref{fig:corr_diff_RG}, we show the correlations of $C_{Hq}^{(1)}$, $C_{Hu}$ and $C_{Hd}$ with selected four-quark operators after including RGE effects in the fit.
As one can see, several correlations are above the $50\%$ level. 
To highlight how much of these correlations is due to the RGE effects, in the top right panel of Fig.~\ref{fig:corr_diff_RG}, we show
the difference between the absolute values between before and after including RGE effects in the fit. 
The absolute correlations of the operators $C_{Hq}^{(1)}$, $C_{Hu}$, and $C_{Hd}$ with $C_{ud}^{(1)}$, $C_{qq}^{(1)}$, $C_{qq}^{(1)\prime}$, $C_{qq}^{(3)}$, $C_{qq}^{(3)\prime}$, and $C_{uu}$ all increase by at least $36\%$. 
The strongest increases for each gauge-quark coefficient are an increase by $92\%$ between $C_{Hq}^{(1)}$ and $C_{qq}^{(1)}$, $82\%$ between $C_{Hu}$ and $C_{qq}^{(1)}$, and $79\%$ between $C_{Hd}$ and $C_{ud}^{(1)}$.
Evidently, the RGE improvement induces strong correlations between the gauge-quark operators and several four-quark operators\footnote{
In our previous SMEFT fit based on partial NLO predictions~\cite{Bartocci:2023nvp}, we noticed an increase of the limits on $C_{Hq}^{(1)}$ by a factor two when including partial NLO EWPO predictions, while the limits on $C_{Hu}$ and $C_{Hd}$ only marginally changed.
As the same diagrams contribute at NLO and in the RGE, this raises the question of why the  correlation effects are stronger in the RGE improved fit than in the partial NLO fit. 
This is likely due to the fact that our NLO fit only includes partial NLO corrections. 
The lack of NLO predictions for $Vh$ production leads to four-quark-operator--independent constraints on operators influencing quark-gauge interactions in the partial NLO fit. 
The oversimplification of the $Zh$-production prediction hence breaks degeneracies between the four-quark operators and $C_{Hq}^{(1)}$, $C_{Hu}$, $C_{Hd}$. For $C_{Hq}^{(1)}$, which has a reduced contribution to $Zh$ production (compare to $C_{Hu}$, $C_{Hd}$ and $C_{Hq}^{(3)}$) as a result of a cancellation between the contributions from up- and down-type quarks, this reduction of correlations with four-quark operators is less pronounced and we can already see its effect in the partial NLO fit. 
This highlights the relevance of further NLO SMEFT predictions, for instance for $Zh$ production, which has recently been calculated for the production at lepton colliders~\cite{Asteriadis:2024xts}.
}.
We have explicitly checked that removing four-quark operators from the fit would reduce the changes of the limits on $C_{Hq}^{(1)}$, $C_{Hu}$, and $C_{Hd}$ in an RGE improved LO fit to the $10\%$ level. Even only removing the four-quark operators $\{C_{uu}, \,C_{dd}, \,  C_{dd}^{\prime}, \, C_{qq}^{(1)}, \, C_{qq}^{(3)} \}$, which are often excluded or constrained to specific directions in top-centered global analyses, already reduces the effect of the RGE improvement significantly, with limits still increasing by factors $2-3$.

The Wilson coefficients of most other operators contributing to EWPO at LO are only mildly influenced by the RGE improvement. The bounds on $C_{HD}$, $C_{Hl}^{(1)}$ and $C_{He}$ increase by $36\%$, $55\%$ and $32\%$, respectively, as a result of increased correlations with various semileptonic operators. Diagonal running effects generally only play a subdominant role for the operators contributing to EWPO at LO. Accounting only for diagonal running effects, the bounds on $C_{HD}$, $C_{Hl}^{(1)}$ and $C_{He}$ would only increase by $16\%$, $13\%$ and $1\%$, respectively.

%----------------------------------------------
%----------------------------------------------
\paragraph{Four-lepton operators}
The four-lepton operators $C_{ll}$, $C_{ee}$, and $C_{le}$ are primarily constrained by lepton scattering processes at LO and exhibit minimal interplay with other datasets. Their RGE effects are suppressed by powers of $\alpha$.
As expected, the limits on these coefficients remain largely unaffected by the inclusion of RGE effects, with global limits showing deviations of less than $7\%$.
The bounds on $C_{ll}^{\prime}$, which enters the SMEFT definition of the Higgs vacuum expectation value and thereby contributes to a larger set of observables, increases by $19\%$ as a result of correlations.

%----------------------------------------------
%----------------------------------------------
\paragraph{Higgs sector}

The limits on most of the Wilson coefficients contributing mainly to Higgs physics do not significantly change after the inclusion of RGE effects. Increases and decreases of the limits in this sector are mostly dominated by diagonal running effects. 
For the coefficient $C_{HG}$ the diagonal running contribution, 
$C_{HG} (\Lambda) \approx 0.75 \, C_{HG}(m_H)$\footnote{Note that we present $C_i(\Lambda)$ as a function of its low scale definition $C_i(\mu)$ here. The proportionality factor defined in this way corresponds more directly to the change of the limit between the LO fit, which sets constraints on $C_i(\mu)$, and the LO+RGE fit, which sets constraints on $C_i(\Lambda)$. If we presented $C_i(\mu)$ as a function of $C_i(\Lambda)$ instead, the limits would have an inverse proportionality.}, 
dominates the decrease of the bound on this coefficient  by around $20\%$.
Similarly, the bound on $C_{H}$ weakens by around $75\%$ dominantly as a result of its diagonal running,
%(in the di-Higgs production prediction), 
$C_H (\Lambda) \approx 1.45 \, C_H (m_H)$.

We also note some shifts of the RGE improved fit limits with respect to the LO fit. As an example, we discuss here the shift of $C_G$ towards more non-SM like values, which is strongly correlated with the shift of $C_{HG}$.
The shift of $C_G$ in the RGE improved fit is rooted in multiple ($<2 \sigma$) deviations in Run~I $t \bar{t}$ production and the fact that it mixes with an operator with a different kinematic structure at the experimentally relevant scale.  
$t \bar{t}$ production is sensitive to modifications of the top-gluon coupling as induced by $C_{\substack{uG\\33}}$, which does not obey a $U(3)^5$ symmetry at the high scale $\Lambda$, but which receives sizable contributions from mixing with $C_G$ at scales $\mu < \Lambda$.

CMS analysed Run~I $t \bar{t}$ data in a double differential distribution in the invariant mass~$m_{t \bar{t}}$ and rapidity~$y_{t \bar{t}}$ of the $t \bar{t}$ system~\cite{CMS:2017iqf,CMS:2013hon}. 
In Fig.~\ref{fig:mtt_pred_shift}, we show the measured $m_{t \bar{t}}$ distribution relative to the SM value. For each bin in $m_{t \bar{t}}$, we show the four measurements in the different rapidity bins. As one can see, all four measurements in the first $m_{t \bar{t}}$ bin lie systematically above the SM prediction, while they lie systematically below the SM prediction in all remaining bins. 
In the same plot, we also show the SMEFT predictions when turning on a single Wilson coefficient $C_G$ (blue solid line) or a single Wilson coefficient $C_{\substack{uG\\33}}$ (orange line) without including RGE effects. 
The values of the Wilson coefficients are chosen to match the RGE improved prediction from $C_G(\Lambda) =1$ at the scale of the first bin $\mu = 340 \text{ GeV}$,
\begin{align}
C_G(340 \text{ GeV}) &= 0.78 \, C_G(4 \text{ TeV}) \, ,
    \nonumber \\
    C_{\substack{uG\\33}}(340 \text{ GeV}) &= -0.13 \, C_G(4 \text{ TeV}) 
    \, .
    \label{eq:CG_ref_values}
\end{align}
Moreover, we show the distribution for $C_G$ when running down to $m_{t\bar{t}}$ from $\Lambda=4 \text{ TeV}$ (dashed blue line). 
Due to a sign difference between the two contributions from $C_G(\Lambda)$ in the SMEFT prediction, they partially cancel each other and lead to a sign flip of the $C_G(\mu)$ prediction of the low-$m_{t \bar{t}}$ bin of the $m_{t \bar{t}}$ kinematic distributions.
As a result, positive values of $C_G$ are able to improve the agreement with the data throughout all measurements when RG effects are included, thereby shifting the $C_G$ limits towards positive values.
\begin{figure}
    \centering
    \includegraphics[height=6.5cm,valign=t]{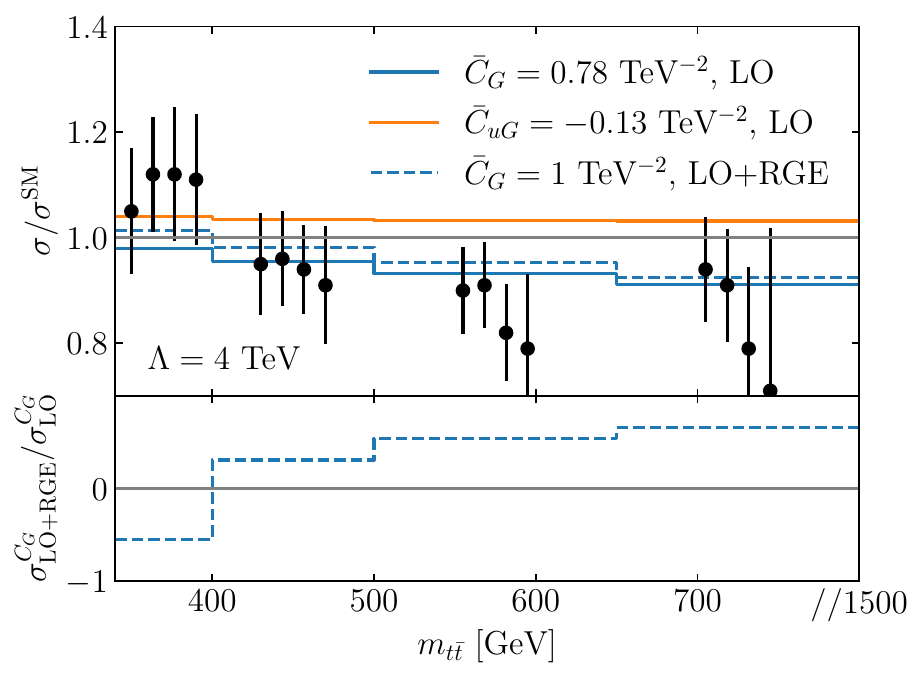}
    \includegraphics[height=6.7cm,valign=t]{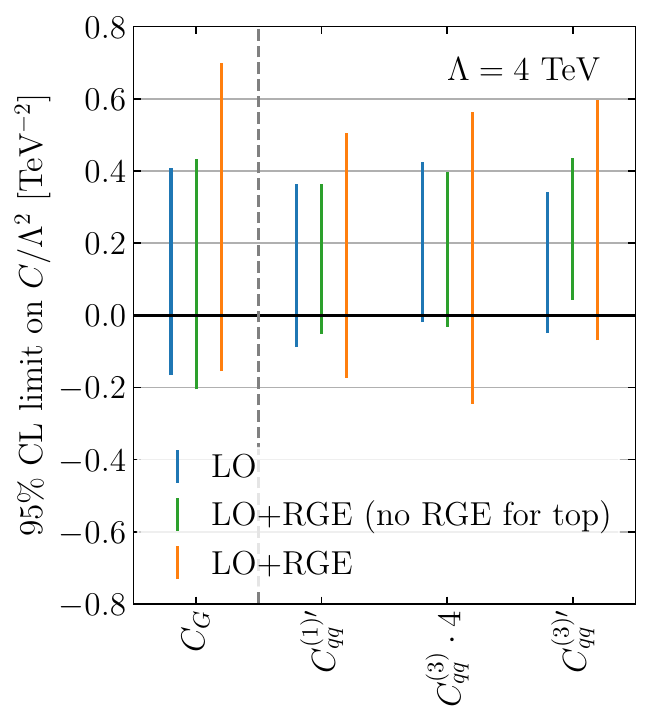}
    \caption{Left: SMEFT predictions for the $m_{t \bar{t}}$ distribution in $t\bar{t}$ production for the operators $C_G$ and $C_{uG}$. We employ the notation $\bar{C} = C/\Lambda^2$. 
    The reference values for $\bar{C}_G$ and $\bar{C}_{uG}$ refer to the RGE-induced contributions from $\bar{C}_G(\Lambda)=1 \text{ TeV}^{-1}$ at the scale $\mu=340 \text{ GeV}$, see Eq.~\eqref{eq:CG_ref_values}.
    The datapoints refer to the CMS Run~I measurements in a double differential $m_{t\bar{t}}-y_{t \bar{t}}$ distribution~\cite{CMS:2017iqf,CMS:2013hon}.
    Right: Limits on selected operators excluding RGE effects in SMEFT predictions for top-quark observables.
    }
    \label{fig:mtt_pred_shift}
\end{figure}

To further highlight that the shift of $C_G$ is dominated by top sector data, we show in Fig.~\ref{fig:mtt_pred_shift} (right) the limits from a global fit when excluding RGE effects in the top sector. 
As apparent from the plot, the limits on these two operators when excluding RGE effects in the top sector (green lines) match the limits from a fit excluding all RGE effects (blue lines) much more closely, and no shift is present. 

Since both $C_{\substack{uG\\33}}$ and $C_{G}$ also enter the predictions of gluon-fusion Higgs production, they are correlated with $C_{HG}$. The shift of $C_G$ in turn leads to a shift of $C_{HG}$.

%----------------------------------------------
%----------------------------------------------
\paragraph{Semileptonic operators}

Semileptonic operators exhibit notable mixing effects with four-quark operators. The most significant impact on the limits is observed for the operator $C_{ed}$, whose bound increases by $45\%$. 
In particular, the operators $C_{dd}$ and $C_{dd}^\prime$, which are among the least constrained in the global fit, mix into $C_{ed}$. Figure~\ref{fig:ced_cdd} illustrates the 95\% CL contours for $C_{ed}$ and $C_{dd}$, marginalising over the remaining 39 parameters. 
When RGE effects are included (orange continuous line), the constraint on $C_{ed}$ is weakened compared to the LO fit without RGE (blue line). However, when the RGE contributions are removed from the DY dataset (orange dashed line) the bound on $C_{ed}$ shrinks back to the blue contour, showing the impact of RGE effects in DY. 
Notably, the correlation between $C_{ed}$ and $C_{dd}$ increases from $-15\%$ to $45\%$ in the global fit after the inclusion of RGE effects.

Other semileptonic operators are less affected by RGE effects. 
The bounds on $C_{eu}$,  $C_{lq}^{(1)}$ and $C_{qe}$ decrease by $25\%$ due to additional constraints from EWPO after the RGE improvement,
which significantly reduces the correlation between selected semileptonic operators. For instance, the correlation between $C_{lq}^{(1)}$ and $C_{ed}$ reduces by $79\%$. 
Meanwhile, the bounds of the remaining coefficients change by less than $15\%$.
\begin{figure}
    \centering
    \includegraphics[width=0.5\linewidth,valign=t]{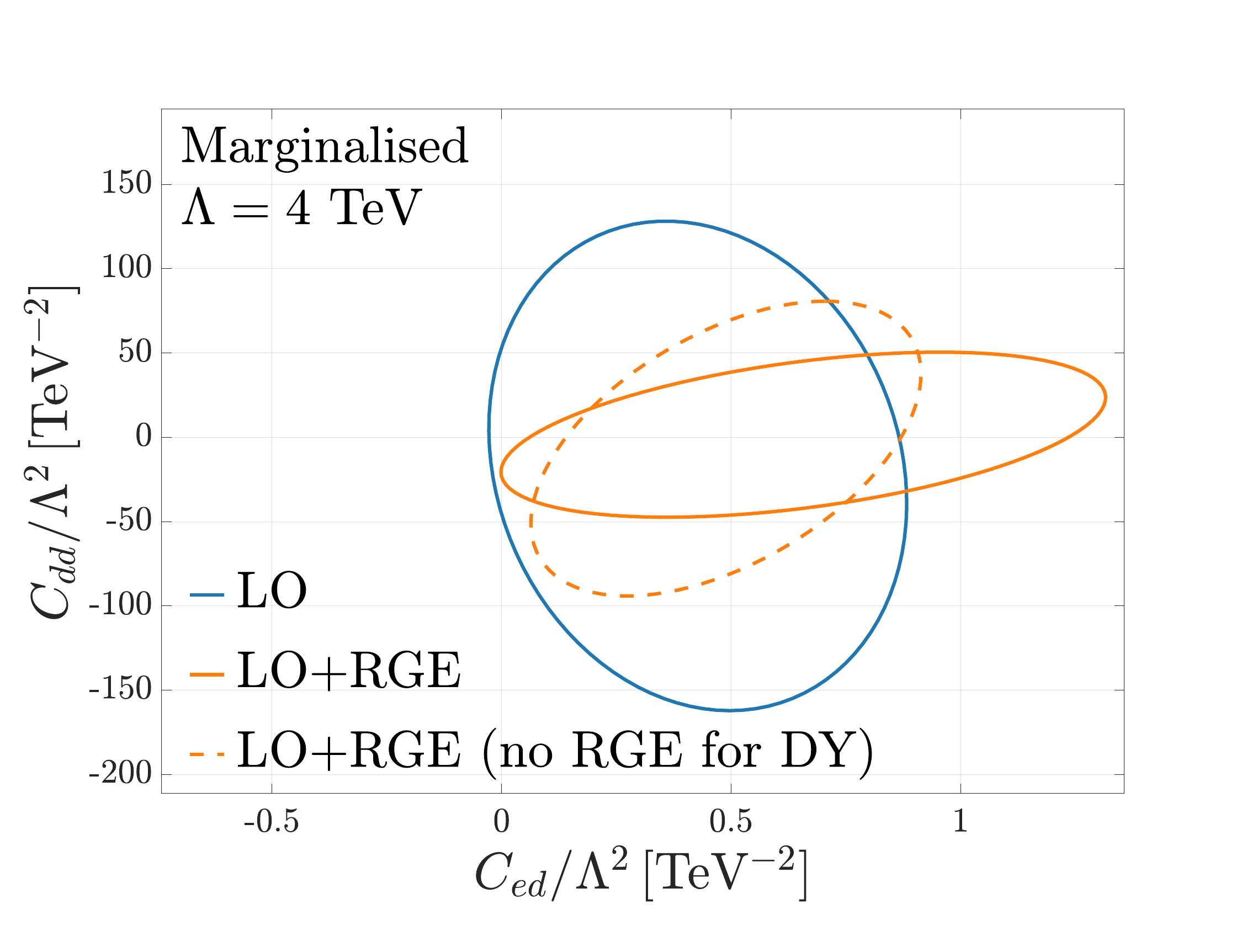}
    \caption{$95\%$ CL limits on the Wilson coefficients $C_{ed}$ and $C_{dd}$ before and after the inclusion of RGE effects in a 2D plane.}
    \label{fig:ced_cdd}
\end{figure}

%----------------------------------------------
%----------------------------------------------
\paragraph{Four-quark operators}
The limits on individual four-quark operators significantly improve in the RG improved LO fit.
As pointed out above, at LO, the Wilson coefficients $C_{qd}^{(1)}$, $C_{qu}^{(1)}$ and $C_{ud}^{(1)}$ are only constrained by the dijets dataset. Since the corresponding operators do not interfere with the dominant SM diagram as a result of their colour structure, the bounds on these operators from an LO fit are weak, leaving (almost) blind directions in the fit. 
Through RGE effects, these operators are constrained in EWPO, PVE and DY, see also the single-parameter fit limits in Fig.~\ref{fig:1D_limits}.
The mixing of $C_{qd}^{(1)}$, $C_{qu}^{(1)}$ and $C_{ud}^{(1)}$ generates sizeable contributions to $C_{Hu}$, $C_{Hd}$ and $C_{qu}^{(1)}$. 
While the strong crosstalk of these and other four-quark coefficients leads to weaker bounds on the coefficients entering EWPO at LO, as discussed above, bounds on the four-quark coefficients in turn improve.
For PVE and DY constraints, the dominant mixing effects are with the semileptonic operator $C_{ed}$. 
While limits from EWPO on $C_{qd}^{(1)}$, $C_{qu}^{(1)}$ and $C_{ud}^{(1)}$ are stronger, they come with large correlations between these four-quark operators. PVE and DY allow to break these degeneracies.

Limits on $C_{qq}^{(1)}$, $C_{uu}$, $C_{dd}$ and $C_{dd}^{\prime}$ improve by $(40-70)\%$. As pointed out already in the discussion of single-parameter limits in Section~\ref{sec:single_parameter}, these coefficients receive additional constraints from DY, EWPO and PVE.

Limits on the coefficients $C_{qq}^{(1)\prime} , C_{qq}^{(3)} , \, C_{qq}^{(3)\prime}$ weaken by $(50-80) \%$ in the RGE improved LO fit.
RGE effects induce significant correlation increases between these four-quark operators, leading to broader bounds on the Wilson coefficients.
These effects are dominated by RGE effects in the top sector, where these operators are tightly constrained, see right panel of Fig.~\ref{fig:mtt_pred_shift}. 
Neglecting RGE effects for the top-sector observables, the limits (green lines) on these three Wilson coefficients  are in close agreement with those of the LO fit (blue lines). 

The bounds on $C_{uu}^{\prime}$ as well as on the octet four-quark operators change by less than $25\%$.

%----------------------------------------------
%----------------------------------------------
\subsection{Partial next-to-leading order predictions}
\label{sec:NLO_effects}
%----------------------------------------------
Finite terms in next-to-leading order (NLO) SMEFT predictions tame some of the degeneracies in the global fit appearing as a result of the RGE improvement. 
As NLO SMEFT predictions are not known for all processes included in our fit, we only include partial NLO SMEFT predictions. Specifically, we include NLO predictions for 
EWPO~\cite{Dawson:2019clf,Dawson:2022bxd,Biekotter:2023xle},
$t\bar{t}h$ production~\cite{Alasfar:2022zyr},
Higgs decays to gluons and photons $h \to gg, \, \gamma \gamma$~\cite{Alasfar:2022zyr},
$t\bar{t}$ production ($m_{t\bar{t}}$ distribution and $t \bar{t}$ charge asymmetries)~\cite{Hartland:2019bjb,Kassabov:2023hbm}.

Formally, the inclusion of NLO SMEFT predictions requires the knowledge of two-loop RG equations. 
As these are currently only partially known~\cite{Alonso:2017tdy,Bern:2020ikv,Jin:2020pwh,Jenkins:2023bls,Fuentes-Martin:2023ljp,DiNoi:2024ajj,Ibarra:2024tpt}, we employ the one-loop RGE as a proxy.

In Fig.~\ref{fig:LO_NLO_RGE}, we present the limits from our RGE improved partial NLO fit (green). Deviations from the RGE improved LO fit can be observed mainly for those operators which are directly impacted by the NLO SMEFT predictions. 
Correlations between $C_{Hq}^{(1)}$, $C_{Hu}$, $C_{Hd}$ and four-quark operators are reduced by NLO SMEFT predictions for EWPO which results in stronger constraints on these Wilson coefficients. 
Limits on some four-quark operators contributing at NLO to $t\bar{t}h$ ($C_{qu}^{(1)}$) and $t\bar{t}$ ($C_{qd}^{(1)}$, $C_{ud}^{(1)}$) production are also reduced when including NLO predictions for these datasets.
The allowed ranges of all other Wilson coefficients are changed by at most $24\%$ ($C_{Hl}^{(3)}$). 

%----------------------------------------------
%----------------------------------------------
\section{Conclusions and outlook}
\label{sec:conclusions}

The inclusion of RGE effects is a crucial advancement for the precision of global SMEFT analyses, allowing for the consistent combination of observables at different energy scales as well as their direct reinterpretation in terms of models defined at high scale. 
We present a global fit of the Wilson coefficients of the $U(3)^5$ symmetric SMEFT to a dataset consisting of observables from EWPO, Higgs, top, flavour, PVE, DY and dijets$+\gamma$ data and including RGE effects within the minimal flavour violation framework.
We find that the limits on most coefficients are only mildly influenced by the inclusion of RGE effects.
However, individual Wilson coefficients may also experience a significant decrease or increase of their bounds as a result of correlations and crosstalk between observables induced through the RGE effects. 

While the impact of the RGE improvement on some Wilson-coefficient bounds is dominated by diagonal running effects, e.g.\ for $C_{HG}$, the limits on others are mostly influenced by their mixing with other coefficients. 
The mixing introduces additional constraints on the coefficients from datasets to which they do not contribute at LO, as well as additional correlations between coefficients. 
A PCA has revealed a significant improvement of the four most weakly constrained directions in our global fit, which were poorly constrained at LO.

The Wilson coefficients corresponding to four-fermion operators are generally improved by the RGE effects, as they mix with operators well constrained in precision measurements. 
This is in particular true for the coefficients of operators with a suppressed interference with the SM at LO, specifically $C_{qd}^{(1)}$, $C_{qu}^{(1)}$ and $C_{ud}^{(1)}$. 
Moreover, the four-down-quark coefficients $C_{dd}$ and $C_{dd}^{\prime}$, which are only weakly constrained at LO, benefit from additional constraints resulting from their mixing with semileptonic operators. 

On the other hand, limits on operators modifying the $Z q \bar{q}(h)$ coupling,
$C_{Hq}^{(1)}, \, C_{Hu}, \, C_{Hd}$, are significantly weakened in an RGE improved fit. 
This is the result of strong correlations between these coefficients and those corresponding to four-quark operators, which are generally much more weakly constrained. 
The modification of limits on operators contributing to EWPO at LO highlights the relevance of the inclusion of observables testing four-quark interactions in global SMEFT fits. 

Finite terms in NLO SMEFT predictions may tame some of the degeneracies induced by RGE effects (as well as those present already at LO) and further improve the bounds in global fits. A conclusive remark on the effects of NLO SMEFT predictions will only be possible when more NLO predictions are available. 

Overall, the inclusion of RGE effects in global SMEFT fits provides a more complete picture of the current status of experimental constraints and facilitates the interpretation in terms of specific models at the NP scale. 
Our study also highlights the interconnectedness of (flavour conserving) four-quark interactions with those tested in precision experiments.  
A generalisation to more generic flavour assumptions at the high scale within the MFV framework is left for future work.

\section*{Acknowledgements}
We thank Matthias K\"onig, Luca Mantani and Ken Mimasu for useful discussions. 
All three authors are supported by  the  Cluster  of  Excellence  ``Precision  Physics,  Fundamental
Interactions, and Structure of Matter" (PRISMA$^+$ EXC 2118/1) funded by the German Research Foundation (DFG) within the German Excellence Strategy (Project ID 390831469). T.H. thanks the CERN theory group for its hospitality during his regular visits to CERN where part of the work was done.

%%%%%%%%%%%%%%%%%%%%%%%%%%%%%%%%%%%%%%%%%%%%%%%%%%%%%%%%%%%%%%%%%%%%%%%%%%%%%%%%%%%%%%%%%%%%%%%%%%%%%%%%%%%%%%%%%%%%%%%%%%%%%%%%%%%%%%%%%%%%%%%%%%%%%%%%%%%%%%%%%%%%%%
%%%%%%%%%%%%%%%%%%%%%%%%%%%%%%%%%%%%%%%%%%%%%%%%%%%%%%%%%%%%%%%%%%%%%%%%%%%%%%%%%%%%%%%%%%%%%%%%%%%%%%%%%%%%%%%%%%%%%%%%%%%%%%%%%%%%%%%%%%%%%%%%%%%%%%%%%%%%%%%%%%%%%%

\appendix
%----------------------------------------------
\section{Principal component analysis}
\label{app:PCA}
\begin{figure}[t]
    \centering
    \includegraphics[width=1.0\textwidth]{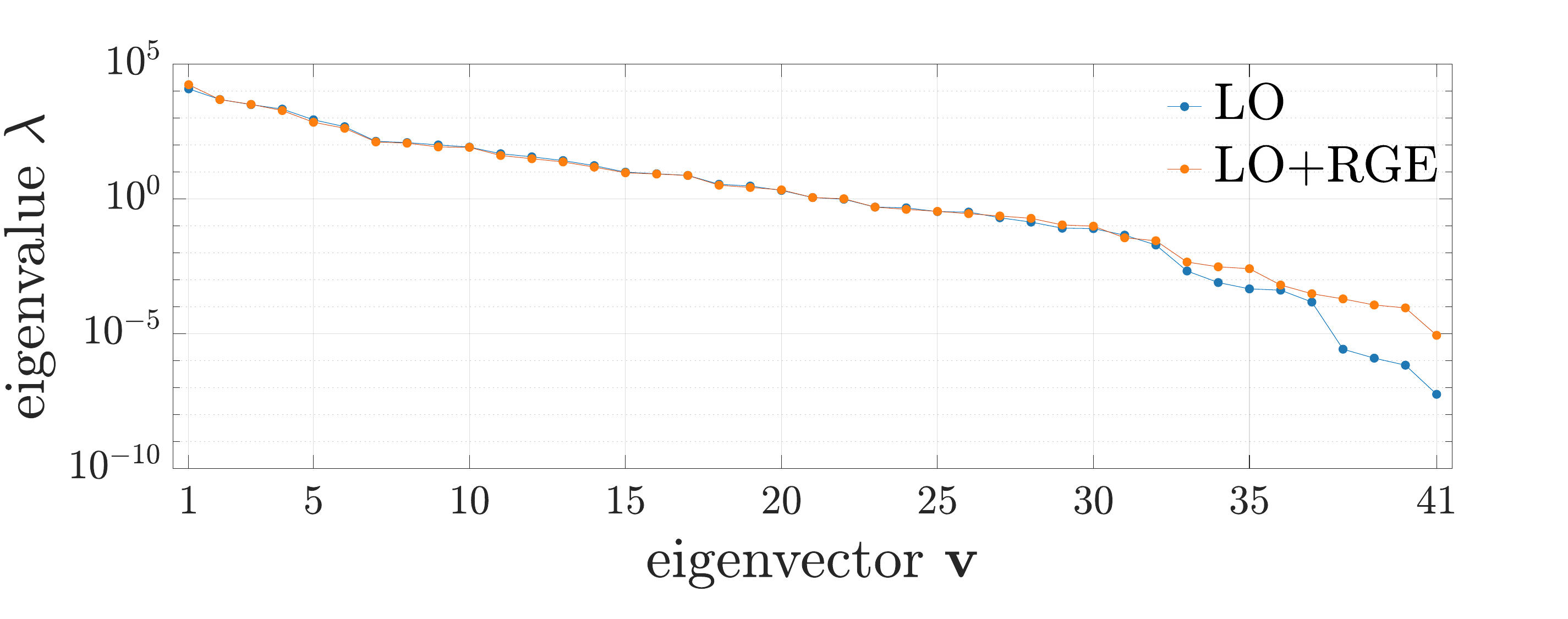}
    \caption{PCA analysis at LO (blue) and LO+RGE (orange). The eigenvalues of the Fisher matrix $\lambda$ are shown for all the 41 eigenvectors $\mathbf{v}$ in decreasing order.}
    \label{fig:PCA}
\end{figure}
In this appendix, we present the results of a principal component analysis (PCA) to examine how the least constrained directions are influenced by the RGE improvement.
The Fisher information matrix, $F(\theta)$, is defined as:
\begin{equation}
F(\theta) = \frac{1}{2} \frac{\partial^2}{\partial C_i \partial C_j} \chi^2 \, ,
\end{equation}
where $\chi^2$ is the chi squared function and $C_i$ denote the Wilson coefficients. 
The eigenvalues $\lambda_i$ of the Fisher matrix provide a measure of how good the directions spanned by the eigenvectors $\mathbf{v}_i$ are constrained in the fit. 
If an eigenvector direction $\mathbf{v}_i$ is unconstrained by the data, its corresponding Fisher matrix eigenvalue will be zero. 

Figure~\ref{fig:PCA} shows the eigenvalues of the Fisher matrix, associated with specific eigenvectors, plotted on a logarithmic scale and ordered from the most to the least constrained direction.
The eigenvalues associated with the most constrained directions are very similar in the fits performed with and without RGE effects. However, the least constrained directions become significantly more constrained when RGE effects are included. 
In particular, we observe a notable reduction in the eigenvalues for the four least constrained directions. Among these, the least constrained direction differs by three orders of magnitude between the two fits, showing how the parameter space results more constrained when RGE is included. 

As we point out in Section~\ref{sec:global}, the limits on individual Wilson coefficients can also be reduced in the RGE improved fit. This mostly concerns the operators $C_{Hu}$, $C_{Hd}$ and $C_{Hq}^{(1)}$. 
The increase of the bounds on these coefficients can also be seen in the PCA, where the coefficients $C_{Hu}$ and $C_{Hd}$ contribute to the seventh least constrained direction in the LO+RGE fit, corresponding to an eigenvalue $\lambda_{35} = 2.6 \times 10^{-3}$. 
In contrast, in the fit without RGE effects, these same coefficients first appear in a direction associated with an eigenvalue $\lambda_{24} = 4.7 \times 10^{-1}$. 
This demonstrates that limits on these coefficients weaken when RGE effects are included.

%----------------------------------------------
\section{Differences with respect to arXiv:2311.04963 }
\label{app:diffs_prev}
%----------------------------------------------

In this appendix, we discuss the influence of differences in the datasets and predictions with respect to Ref.~\cite{Bartocci:2023nvp} on the global fit. 
We show in Figure \ref{fig:previous_vs_present_fit} the comparison between the previous paper and the present work\footnote{Factors of two in the normalisation of the Wilson coefficients of some four-quark operators, as discussed in Section~\ref{sec:SMEFT_pred} have been taken into account.}, reporting those coefficients where the bounds changed on at least one side by more than $15\%$.

\begin{figure}[t]
    \centering
    \includegraphics[width=0.65\textwidth]{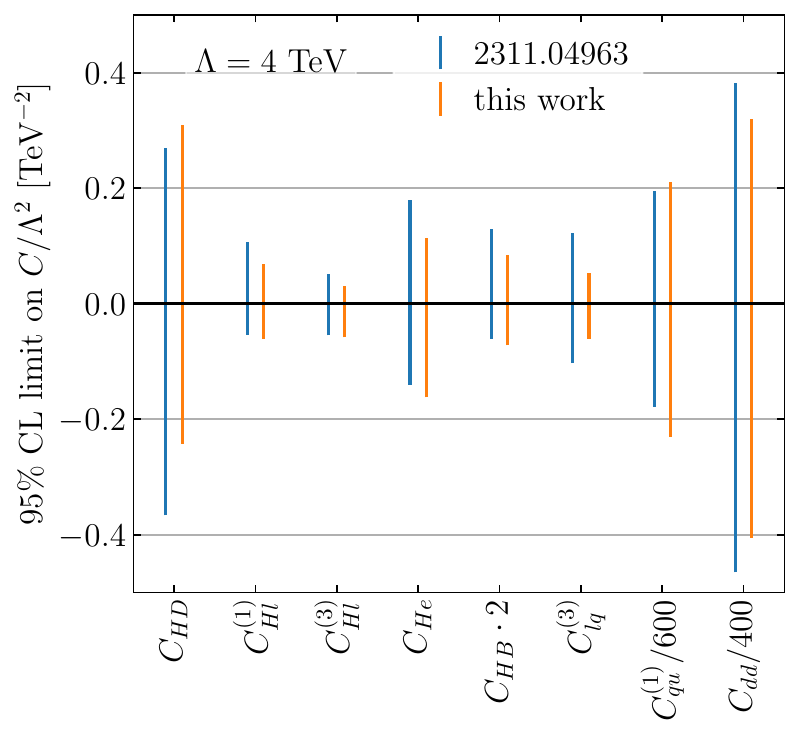}
    \caption{LO global analysis of all 41 Wilson coefficients using the dataset of Ref.~\cite{Bartocci:2023nvp} and using the dataset of the present paper. Only coefficients for which the bounds change by more than $15\%$ on at least one side are shown.}
    \label{fig:previous_vs_present_fit}
\end{figure}
\enlargethispage{5mm}
%------------------------------
\subsection*{Inclusion of $\Delta_{\text{CKM}}$ and diboson data}
With respect to Ref.~\cite{Bartocci:2023nvp}, we have added a pseudo-observable for $\Delta_{\text{CKM}}$ as well as two kinematic distributions of $WZ$~\cite{ATLAS:2019bsc} and $WW$~\cite{ATLAS:2019rob} production. 
Observables from $\beta$-decay and semileptonic meson decays~\cite{Gonzalez-Alonso:2016etj,Falkowski:2020pma}
have been shown to have an interesting interplay with EWPO~\cite{Cirigliano:2022qdm,ThomasArun:2023wbd,Cirigliano:2023nol}. 
As it was already done in the Appendix B of \cite{Bartocci:2023nvp}, we include these observables as a single pseudo-observable $\Delta_{\text{CKM}}$ representing the unitarity of the CKM matrix. This inclusion has a sizeable impact on $C_{lq}^{(3)}$ only.

Just like Higgs data, LHC diboson observables are known to break degeneracies present in EWPO. 
The inclusion of the diboson datasets reduces the allowed Wilson coefficient ranges of the operators contributing to EWPO at LO by up to $23\%$. The limits on the correlated Wilson coefficients $C_{HB}$ improves at a similar level.

%------------------------------
\subsection*{EWPO predictions}
In Ref.~\cite{Bartocci:2023nvp}, we used EWPO SMEFT predictions from~\cite{Dawson:2019clf} which employs the on-shell definition $\alpha^{\text{O.S.}}(0)$ as an input parameter. Here, we use the $\overline{\text{MS}}$ definition of $\alpha$ in a five-flavour version of QED×QCD as an input instead, and use the SMEFT predictions from~\cite{Biekotter:2023xle}. This is more consistent with the input parameter choice for other observables included in our fit. 
Using a different renormalisation scheme for $\alpha$ slightly modifies some prefactors of the EWPO SMEFT predictions.
Using the updated EWPO SMEFT predictions in our global fit, we do not find effects above $15\%$.
\subsection*{Dijets$+\gamma$ predictions}
\label{app:dijet_plus_photon}
In Ref.~\cite{Bartocci:2023nvp}, we used predictions for ATLAS dijet plus photon production~\cite{ATLAS:2019itm} determined with \texttt{SMEFTsim} \cite{Brivio:2020onw} under the $U(3)^5$-model assumption. In this paper, due to the need of explicit flavour indices in order to correctly include RGE effects, we generated predictions under the general flavour assumption (model \texttt{SMEFTsim\_general\_alphaScheme\_UFO}).
We provide these new predictions as an ancillary file with the arXiv submission. 
We compare these new predictions, after restoring the flavour symmetry, with the previous ones and find an average deviation for each coefficient below $3\%$. In the global fit, this introduces some $15\%$ deviations for the four-quark operators $C_{qu}^{(1)}$ and $C_{dd}$.
%----------------------------------------------

\FloatBarrier

%----------------------------------------------
\newpage
\section{Flavour symmetric and CP even operators}
\label{app:operators}
%----------------------------------------------

\begin{table}[h]
\begin{center}
\small
\begin{minipage}[t]{4.4cm}
\renewcommand{\arraystretch}{1.5}
\begin{tabular}[t]{c|c}
\multicolumn{2}{c}{$1:X^3$} \\
\hline
$Q_G$                & $f^{ABC} G_\mu^{A\nu} G_\nu^{B\rho} G_\rho^{C\mu} $ \\
$Q_W$                & $\epsilon^{IJK} W_\mu^{I\nu} W_\nu^{J\rho} W_\rho^{K\mu}$ \\ 
\end{tabular}
\end{minipage}
\begin{minipage}[t]{2.6cm}
\renewcommand{\arraystretch}{1.5}
\begin{tabular}[t]{c|c}
\multicolumn{2}{c}{$2:H^6$} \\
\hline
$Q_H$       & $(H^\dag H)^3$ 
\end{tabular}
\end{minipage}
\begin{minipage}[t]{5.1cm}
\renewcommand{\arraystretch}{1.5}
\begin{tabular}[t]{c|c}
\multicolumn{2}{c}{$3:H^4 D^2$} \\
\hline
$Q_{H\Box}$ & $(H^\dag H)\Box(H^\dag H)$ \\
$Q_{H D}$   & $\ \left(H^\dag D_\mu H\right)^* \left(H^\dag D_\mu H\right)$ 
\end{tabular}
\end{minipage}

\vspace{0.25cm}

\begin{minipage}[t]{4.6cm}
\renewcommand{\arraystretch}{1.5}
\begin{tabular}[t]{c|c}
\multicolumn{2}{c}{$4:X^2H^2$} \\
\hline
$Q_{H G}$     & $H^\dag H\, G^A_{\mu\nu} G^{A\mu\nu}$ \\
$Q_{H W}$     & $H^\dag H\, W^I_{\mu\nu} W^{I\mu\nu}$ \\
$Q_{H B}$     & $ H^\dag H\, B_{\mu\nu} B^{\mu\nu}$ \\
$Q_{H WB}$     & $ H^\dag \tau^I H\, W^I_{\mu\nu} B^{\mu\nu}$ \\
\end{tabular}
\end{minipage}
\begin{minipage}[t]{5.2cm}
\renewcommand{\arraystretch}{1.5}
\begin{tabular}[t]{c|c}
\multicolumn{2}{c}{$7:\psi^2H^2 D$} \\
\hline
$Q_{H l}^{(1)}$      & $(H^\dag i\overleftrightarrow{D}_\mu H)(\bar l_p \gamma^\mu l_p)$\\
$Q_{H l}^{(3)}$      & $(H^\dag i\overleftrightarrow{D}^I_\mu H)(\bar l_p \tau^I \gamma^\mu l_p)$\\
$Q_{H e}$            & $(H^\dag i\overleftrightarrow{D}_\mu H)(\bar e_p \gamma^\mu e_p)$\\
$Q_{H q}^{(1)}$      & $(H^\dag i\overleftrightarrow{D}_\mu H)(\bar q_p \gamma^\mu q_p)$\\
$Q_{H q}^{(3)}$      & $(H^\dag i\overleftrightarrow{D}^I_\mu H)(\bar q_p \tau^I \gamma^\mu q_p)$\\
$Q_{H u}$            & $(H^\dag i\overleftrightarrow{D}_\mu H)(\bar u_p \gamma^\mu u_p)$\\
$Q_{H d}$            & $(H^\dag i\overleftrightarrow{D}_\mu H)(\bar d_p \gamma^\mu d_p)$\\
\end{tabular}
\end{minipage}

\vspace{0.25cm}

\begin{minipage}[t]{4.75cm}
\renewcommand{\arraystretch}{1.5}
\begin{tabular}[t]{c|c}
\multicolumn{2}{c}{$8:(\bar LL)(\bar LL)$} \\
\hline
$Q_{\ell \ell}$        & $(\bar l_p \gamma_\mu l_p)(\bar l_s \gamma^\mu l_s)$ \\
$Q_{\ell \ell}^\prime$        & $(\bar l_p \gamma_\mu l_s)(\bar l_s \gamma^\mu l_p)$ \\
$Q_{qq}^{(1)}$  & $(\bar q_p \gamma_\mu q_p)(\bar q_s \gamma^\mu q_s)$ \\
$Q_{qq}^{(3)}$  & $(\bar q_p \gamma_\mu \tau^I q_p)(\bar q_s \gamma^\mu \tau^I q_s)$ \\
$Q_{qq}^{(1)\prime}$  & $(\bar q_p \gamma_\mu q_s)(\bar q_s \gamma^\mu q_p)$ \\
$Q_{qq}^{(3)\prime}$  & $(\bar q_p \gamma_\mu \tau^I q_s)(\bar q_s \gamma^\mu \tau^I q_p)$ \\
$Q_{\ell q}^{(1)}$                & $(\bar l_p \gamma_\mu l_p)(\bar q_s \gamma^\mu q_s)$ \\
$Q_{\ell q}^{(3)}$                & $(\bar l_p \gamma_\mu \tau^I l_p)(\bar q_s \gamma^\mu \tau^I q_s)$ 
\end{tabular}
\end{minipage}
\begin{minipage}[t]{5.25cm}
\renewcommand{\arraystretch}{1.5}
\begin{tabular}[t]{c|c}
\multicolumn{2}{c}{$8:(\bar RR)(\bar RR)$} \\
\hline
$Q_{ee}$               & $(\bar e_p \gamma_\mu e_p)(\bar e_s \gamma^\mu e_s)$ \\
$Q_{uu}$        & $(\bar u_p \gamma_\mu u_p)(\bar u_s \gamma^\mu u_s)$ \\
$Q_{uu}^\prime$        & $(\bar u_p \gamma_\mu u_s)(\bar u_s \gamma^\mu u_p)$ \\
$Q_{dd}$        & $(\bar d_p \gamma_\mu d_p)(\bar d_s \gamma^\mu d_s)$ \\
$Q_{dd}^\prime$        & $(\bar d_p \gamma_\mu d_s)(\bar d_s \gamma^\mu d_p)$ \\
$Q_{eu}$                      & $(\bar e_p \gamma_\mu e_p)(\bar u_s \gamma^\mu u_s)$ \\
$Q_{ed}$                      & $(\bar e_p \gamma_\mu e_p)(\bar d_s\gamma^\mu d_s)$ \\
$Q_{ud}^{(1)}$                & $(\bar u_p \gamma_\mu u_p)(\bar d_s \gamma^\mu d_s)$ \\
$Q_{ud}^{(8)}$                & $(\bar u_p \gamma_\mu T^A u_p)(\bar d_s \gamma^\mu T^A d_s)$ \\
\end{tabular}
\end{minipage}
\begin{minipage}[t]{4.75cm}
\renewcommand{\arraystretch}{1.5}
\begin{tabular}[t]{c|c}
\multicolumn{2}{c}{$8:(\bar LL)(\bar RR)$} \\
\hline
$Q_{le}$               & $(\bar l_p \gamma_\mu l_p)(\bar e_s \gamma^\mu e_s)$ \\
$Q_{lu}$               & $(\bar l_p \gamma_\mu l_p)(\bar u_s \gamma^\mu u_s)$ \\
$Q_{ld}$               & $(\bar l_p \gamma_\mu l_p)(\bar d_s \gamma^\mu d_s)$ \\
$Q_{qe}$               & $(\bar q_p \gamma_\mu q_p)(\bar e_s \gamma^\mu e_s)$ \\
$Q_{qu}^{(1)}$         & $(\bar q_p \gamma_\mu q_p)(\bar u_s \gamma^\mu u_s)$ \\ 
$Q_{qu}^{(8)}$         & $(\bar q_p \gamma_\mu T^A q_p)(\bar u_s \gamma^\mu T^A u_s)$ \\ 
$Q_{qd}^{(1)}$ & $(\bar q_p \gamma_\mu q_p)(\bar d_s \gamma^\mu d_s)$ \\
$Q_{qd}^{(8)}$ & $(\bar q_p \gamma_\mu T^A q_p)(\bar d_s \gamma^\mu T^A d_s)$\\
\end{tabular}
\end{minipage}
\end{center}
\caption{\label{tab:basis}
Flavour symmetric and CP even dimension-six SMEFT operators in the Warsaw basis. }
\end{table}

\newpage
%----------------------------------------------
\section{Observables}
\label{app:observables}
%----------------------------------------------
We list the observables included in our analysis in Tables~\ref{tab:obset}-\ref{tab:obset_PVE_flavour}.
Table~\ref{tab:obset} lists observables from Higgs physics, LHC diboson and $Zjj$~production, Tables~\ref{tab:obset_top} and~\ref{tab:obset_top2_DY_dijet} list observables from the top sector as well as Drell-Yan and dijet+photon data, and Table~\ref{tab:obset_PVE_flavour} lists observables from EWPO, PVE, lepton scattering and flavour.

\begin{table}[h]
	\centering
	\renewcommand{\arraystretch}{2.0}
    \caption{Higgs and electroweak observables included in the fit.}
	\begin{adjustbox}{width=0.9\textwidth}
		\label{tab:obset}
		\begin{tabular}{|c|c|c|c|c|}
			\hline
			\multicolumn{2}{|c|}{Observables} & no. of measurements	 & scale(s) &  References \\
			\hline
			\multicolumn{2}{|c|}{\bf{Higgs Data}} & 159 & & \multirow{1}{*}{}  \\ 
			\cline{1-5}
			\multirow{3}{*}{7 and 8 TeV } & ATLAS \& CMS combination  & \multirow{1}{*}{20} & $M_H$ & \multirow{1}{*}{Table~8 of Ref.~\cite{Khachatryan:2016vau}} \\
			\cline{2-5}
			\multirow{3}{*}{Run-I data }& ATLAS \& CMS combination $\mu( h \to \mu \mu)$ &  \multirow{1}{*}{1} &  $M_H$ &  \multirow{1}{*}{Table~13 of Ref.~\cite{Khachatryan:2016vau}}  \\ 
            \cline{2-5}
			& ATLAS $\mu (h \to Z \gamma)$ & \multirow{1}{*}{1} & $M_H$ &  \multirow{1}{*}{ Figure~1 of Ref.~\cite{Aad:2015gba}}  \\
			\hline
			\multirow{4}{*}{13 TeV ATLAS} &  $\mu ( h \to Z \gamma )$ at 139 $\ifb$ & 1 & $M_H$ &  \cite{Aad:2020plj} \\
			& $\mu ( h \to \mu \mu)$ at 139 $\ifb$ & 1 & $M_H$ &  \cite{Aad:2020xfq} \\
			Run-II data  & $\mu(h \to \tau \tau)$ at 139 $\ifb$ & 4 & $M_H$ & Figure~14 of Ref.~\cite{ATLAS-CONF-2021-044} \\
		    & $\mu( h \to bb)$ in VBF and ${ttH}$ at 139 $\ifb$ & 1+1 & $M_H$  & \cite{ATLAS:2020bhl,ATLAS:2020syy}  \\
		    \cline{2-5} 
		    & STXS $h \to \gamma \gamma/ZZ/b \bar{b}$ at $139\ifb$ & 42 & $\sqrt{M_H^2 + (p_T^H)^2}$ & Figures~1 and 2 of Ref.~\cite{ATLAS:2020naq} \\
			& STXS $ h \rightarrow$ $W W$ in ggF, VBF at $139\ifb$ & 11 & $\sqrt{M_H^2 + (p_T^H)^2}$ & Figures~12 and 14 of Ref.~\cite{ATLAS:2021upe} \\
            \cline{2-5}
            & di-Higgs $\mu_{_{HH}}^{b\bar{b}b \bar{b}}$, $\mu_{_{HH}}^{b \bar{b}\tau \bar{\tau}}$, $\mu_{_{HH}}^{b \bar{b} \gamma \gamma}$ & 3 & $M_H$  &   \cite{ATLAS:2018dpp,ATLAS:2018rnh,ATLAS:2018uni}  \\
			\hline
			 &  $\mu(h \to b \bar{b})$ in $Vh$ at $35.9/41.5\ifb$ & 2  &  $M_H$  & Table~4 of Ref.~\cite{CMS:2020gsy} \\
			 &  $\mu(h \to W W)$ in ggF at $137\ifb$ & 1  & $M_H$ & \cite{CMS:2020dvg} \\
			 13 TeV CMS &  $\mu (h\to \mu \mu)$ at  $137\ifb$ & 4  & $M_H$ & Figure~11 of Ref.~\cite{CMS:2020xwi} \\
			Run-II data & $\mu (h \to \tau \tau/WW)$ in  $t\bar{t}h$ at $137\ifb$ & 3  & $M_H$ & Figure~14 of Ref.~\cite{CMS:2020mpn} \\
			\cline{2-5} 
			 & STXS $h\to WW$ at $137\ifb$ in $Vh$  & 4 & $\sqrt{M_H^2 + (p_T^H)^2}$ &  Table~9 of Ref.~\cite{CMS:2021ixs} 
			\\
			& STXS $h \to \tau \tau$ at  $137\ifb$  & 11 & $\sqrt{M_H^2 + (p_T^H)^2}$ &  Figures~11/12 of Ref.~\cite{CMS:2020dvp} 
			\\
			& STXS $h \to \gamma \gamma$ at  $137\ifb$ & 27 &$\sqrt{M_H^2 + (p_T^H)^2}$ & Table~13 and Figure~21 of Ref.~\cite{CMS:2021kom} 
			\\
			& STXS $h \to ZZ $ at  $137\ifb$ & 18 & $\sqrt{M_H^2 + (p_T^H)^2}$ & Table~6 and Figure~15 of Ref.~\cite{CMS:2021ugl} 
			\\
            \cline{2-5}
            & di-Higgs $\mu_{_{HH}}^{b\bar{b}b \bar{b}}$, $\mu_{_{HH}}^{b \bar{b}\tau \bar{\tau}}$, $\mu_{_{HH}}^{b \bar{b} \gamma \gamma}$ & 3  & $M_H$ &   \cite{CMS:2020tkr, CMS:2021ssj,CMS:2017hea}  \\
			% %
			\hline
			\multicolumn{2}{|c|}{{\bf{ ATLAS $Zjj$  13 TeV $\Delta \phi_{jj}$}} at $139\ifb$}&12  & $M_Z$  & Figure~7(d) of Ref.~\cite{ATLAS:2020nzk}  \\
            \hline
			\multicolumn{2}{|c|}{{\bf{ ATLAS $WZ$  13 TeV $p_{T}^{Z}$}} at $36.1\ifb$}& 7  & $\sqrt{M_Z^2 + (p_T^{Z})^2}$  & Figure~4(a) of Ref.~\cite{ATLAS:2019bsc}  \\
            \hline
			\multicolumn{2}{|c|}{{\bf{ ATLAS $WW$  13 TeV $p_T^{\ell, \text{lead}}$}} at $36.1\ifb$}&14  & $\sqrt{M_W^2 + (p_T^{\ell, \text{lead}})^2 }$  & Figure~7(a) of Ref.~\cite{ATLAS:2019rob}  \\
            \hline
		\end{tabular}
	\end{adjustbox}
\end{table}

\begin{table}[ht]
	\centering
	\renewcommand{\arraystretch}{2.0}
	\caption{Top physics observables from Tevatron and LHC Run I included in the fit.}
	\begin{adjustbox}{width=0.9\textwidth}
		\label{tab:obset_top}
		\begin{tabular}{|c|c|c|c|c|}
			\hline
			\multicolumn{2}{|c|}{Observables} & no. of meas. & scale(s) &  References \\
			\hline
        \multicolumn{2}{|c|}{\bf{Top Data from Tevatron and LHC Run I}} & 82 &  & \\ 
			\cline{1-5} 
 Tevatron & %combination of 
 forward-backward asymmetry $A_{FB}(m_{t\bar{t}})$ for $\mathrm{t}\overline{\mathrm{t}}$ production  &
$4$ & $\max ( m_t , \, m_{t\bar{t}} )$ &
 ~\cite{CDF:2017cvy} \\ \hline
\multirow{2}{*}{ATLAS \& CMS}
    & %combination of 
    charge asymmetry $A_C(m_{t\bar{t}})$ for $\mathrm{t}\overline{\mathrm{t}}$ production in the $\ell$+jets channel  &
 $6$ & $\max ( m_t , \, m_{t\bar{t}} )$ &
 ~\cite{ATLAS:2017gkv} \\ \cline{2-5} 
 & %combination of 
   $W$-boson helicity fractions in top decay %. $f_0,\, f_L\,\& \, f_R$
 & $3$ & $m_t$ &  ~\cite{CMS:2020ezf} \\ 
  \hline
 \multirow{8}{*}{ATLAS} &  charge asymmetry $A_C(m_{t\bar{t}})$ for $\mathrm{t}\overline{\mathrm{t}}$ production in the dilepton channel  &
 $1$ & $m_t$ &
 ~\cite{ATLAS:2016ykb} \\  \cline{2-5} 
     & $\sigma_{t\bar{t}W}, \, \sigma_{t\bar{t}Z}$ &
 $2$ & $m_t$ &
 ~\cite{ATLAS:2015qtq} \\  \cline{2-5} 
   & $\tfrac{d\sigma}{dp^T_{t}} ,\quad \tfrac{d\sigma}{d|y_{\bar{t}}|}$  for $t$-channel single-top production
 &
 $4+5$ & $\sqrt{m_t + (p_T^t)^2}$, $m_t$&
 ~\cite{ATLAS:2017rso} \\ \cline{2-5} 
  &  $\sigma_{tW}$ in the single lepton channel &
 $1$ & $m_t$ & ~\cite{ATLAS:2020cwj} \\ \cline{2-5} 
 & $\sigma_{tW}$ in the dilepton channel &
 $1$ & $m_t$ & ~\cite{ATLAS:2015igu} \\ \cline{2-5} 
& $s$-channel single-top cross section &
 $1$ & $m_t$ & ~\cite{ATLAS:2015jmq} \\ \cline{2-5} 
& $\tfrac{d\sigma}{dm_{t\bar{t}}}$ for $t\bar{t}$ production in the dilepton channel
  &
 $6$ & $m_{t\bar{t}}$   &
 ~\cite{ATLAS:2016pal} \\ \cline{2-5} 
 & $\tfrac{d\sigma}{dp^T_{t}}$  for $t\bar{t}$ production in the $\ell$+jets channel
 & $8$ &  $\sqrt{m_t + (p_T^t)^2}$  &
 ~\cite{ATLAS:2015lsn} \\ \hline
  \multirow{10}{*}{CMS}    & $\sigma_{t\bar{t}\gamma}$ in the $\ell+$ jets channel&
 $1$ &  $m_t$  &
 ~\cite{CMS:2015uvn} \\ \cline{2-5} 
  & charge asymmetry $A_C(m_{t\bar{t}})$ for $\mathrm{t}\overline{\mathrm{t}}$ production in the dilepton channel  &
 $3$ & $\max ( m_t , \, m_{t\bar{t}} )$   &
 ~\cite{CMS:2016ypc} \\ \cline{2-5} 
  & $\sigma_{t\bar{t}W}, \, \sigma_{t\bar{t}Z}$ &
 $2$ &  $m_t$  &
  ~\cite{CMS:2015uvn} \\ \cline{2-5} 
   & $\sigma_{t\bar{t}\gamma}$ in the $\ell+$ jets channel.&
 $1$ &  $m_t$  &
  ~\cite{CMS:2017tzb} \\ \cline{2-5} 
   & $s$-channel single-top cross section &
 $1$ &  $m_t$  &
 ~\cite{CMS:2016xoq} \\ \cline{2-5} 
  &   $\tfrac{d\sigma}{dp_T^{t+\bar{t}}}$ of $t$-channel single-top production
 &
 $6$ &  $\sqrt{m_t^2 + (p_T^t)^2}$  & ~\cite{CMS:2014ika} \\ \cline{2-5} 
  & $t$-channel single-top and anti-top cross sections $R_t$ &
 $1$ &  $m_t$  & ~\cite{CMS:2014mgj} \\ \cline{2-5} 
  & $\sigma_{tW}$ &
 $1$ &  $m_t$  & ~\cite{CMS:2014fut} \\ \cline{2-5} 
 %%%%%%%%%%%%%%%%%%
   & $\tfrac{d\sigma}{dm_{t\bar{t}}dy_{t\bar{t}}}$ for $t\bar{t}$ production in the dilepton channel
  & $16$ &  $m_{t\bar{t}}$   &
 ~\cite{CMS:2017iqf,CMS:2013hon} \\ \cline{2-5} 
  &  $\tfrac{d\sigma}{dp^T_{t}}$ for $t\bar{t}$ production in the $\ell$+jets channel
  & $8$ & $\sqrt{m_t^2 + (p_T^t)^2}$   &
 ~\cite{CMS:2015rld,CMS:2016csa} \\ \hline
		\end{tabular}
	\end{adjustbox}
\end{table}
 
\begin{table}[ht]
	\centering
	\renewcommand{\arraystretch}{2.0}
	\caption{Top physics observables from LHC Run~II as well as data from Drell-Yan and dijet+photon production included in the analysis. }
	\begin{adjustbox}{width=0.9\textwidth}
		\label{tab:obset_top2_DY_dijet}
		\begin{tabular}{|c|c|c|c|c|}
			\hline
			\multicolumn{2}{|c|}{Observables} & no. of meas. & scale(s) &  References \\
			\hline
        \multicolumn{2}{|c|}{\bf{Top Data from LHC Run II}} & 55 & & \\ 
			\cline{1-5} 
 %%%%%
  \multirow{6}{*}{ATLAS} & $\sigma_{tW}$  &
 $1$ &  $m_t$  &
 ~\cite{ATLAS:2016ofl} \\ \cline{2-5} 
   & $\sigma_{tZ}$ &
 $1$ &  $m_t$  &
 ~\cite{ATLAS:2017dsm} \\ \cline{2-5} 
  & $\sigma_{t+\bar{t}}, \, R_t$ for $t$-channel single-top and anti-top cross sections&
 1+1 &  $m_t$  &~\cite{ATLAS:2016qhd} \\ \cline{2-5} 
   &   charge asymmetry $A_C(m_{t\bar{t}})$ for $\mathrm{t}\overline{\mathrm{t}}$ production &
 $5$ &  $\max ( m_t , \, m_{t\bar{t}} )$  &
 ~\cite{ATLAS:2019czt} \\ \cline{2-5} 
  &  $\sigma_{t\bar{t}W}, \, \sigma_{t\bar{t}Z}$  &
 $2$ &  $m_t$  &
 ~\cite{ATLAS:2019fwo} \\ \cline{2-5} 
   &  $\tfrac{d\sigma}{dp_T^{\gamma}}$ for  $t\bar{t}\gamma$ production
 &
 $11$ &   $\sqrt{m_t^2+ (p_T^\gamma)^2}$   &
 ~\cite{ATLAS:2020yrp} \\ \hline
  \multirow{7}{*}{CMS} & $\sigma_{tW}$ &
 $1$ &  $m_t$  &
 ~\cite{CMS:2018amb} \\ \cline{2-5} 
   & $\sigma_{tZ}$ in the $Z\to\ell^+\ell^-$ channel  &
 $1$ &  $m_t$  &
 ~\cite{CMS:2018sgc} \\ \cline{2-5} 
 & %differential cross sections and charge ratios 
 $\tfrac{d\sigma}{dp_T^{t+\bar{t}}}$ and $R_t\left(p_T^{t+\bar{t}}\right)$ for $t$-channel single-top quark production &
 $5+5$ & $\sqrt{m_t^2 + (p_T^t)^2}$   &
 ~\cite{CMS:2019jjp} \\ \cline{2-5} 
   & $\tfrac{d\sigma}{dm_{t\bar{t}}}$  for $t\bar{t}$ production in the dilepton channel
 &
 $6$ &  $m_{t\bar{t}}$  &
 ~\cite{CMS:2018fks}  \\ \cline{2-5} 
 & $\tfrac{d\sigma}{dm_{t\bar{t}}}$ for $t\bar{t}$ production in the $\ell+$jets channel
&
 $15$ &  $m_{t\bar{t}}$  &
 \cite{CMS:2021fhl}  \\ \cline{2-5} 
 & $\sigma_{t\bar{t}W}$ &  
 $1$ & $m_t$   &
 ~\cite{CMS:2017ugv} \\ \cline{2-5} 
 & $\tfrac{d\sigma}{dp_T^{Z}}$ for $t\bar{t}Z$ production  &
 $4$ &  $\sqrt{m_t^2+(p_T^Z)^2}$  &
 ~\cite{CMS:2019too} \\ 
 \hline
     			\multicolumn{2}{|c|}{\bf{Drell-Yan}} & 109 & &  \\ 
			\cline{1-5}
			\multirow{3}{*}{13 TeV } & CMS $e^+ e^-$, $m_{ee}$ & 61 (up to 3~TeV) & $m_{ee}$& Figure~2 of~\cite{CMS:2021ctt}  \\
			\cline{2-5}
			& CMS $\mu^+ \mu^-$, $m_{\mu \mu}$&  34 (up to 3~TeV) & $m_{\mu\mu}$&  Figure~2 of~\cite{CMS:2021ctt} \\ 
            \cline{2-5}
			& ATLAS $\tau^+\tau^-$, $m_T^{\text{tot}}$ & 14 (up to 3~TeV) & $m_T^{\text{tot}}$ & Figure~1 of~\cite{ATLAS:2020zms} \\
   \hline
			\multicolumn{2}{|c|}{\bf{Dijets+photon}} & 26 & & \\ 
			\cline{1-5}
			\multirow{1}{*}{13 TeV } & ATLAS $\frac{d N_{\text{evt}} }{dm_{jj}}$ for $pp\to jj\gamma+X$& 26 (from 500~GeV) & $m_{jj}$ & Figure~1 of~\cite{ATLAS:2019itm}  \\
 \hline
		\end{tabular}
	\end{adjustbox}
\end{table}

\begin{table}[h!]
	\centering
    \vspace{-1cm}
	\renewcommand{\arraystretch}{2.0}
    \caption{EWPO, PVE, lepton scattering and flavour observables included in the fit. For a summary of the experimental input values of the EWPO observables used, see also Tab.~3 of~\cite{Dawson:2019clf}. $^*$ For scales below $5$ GeV, we only include the LEFT RGE up to this scale. We explicitly checked that the evolution below $5$ GeV is almost negligible when compared with the evolution between $\Lambda$ and $m_b$.}
	\begin{adjustbox}{width=0.9\textwidth}
		\label{tab:obset_PVE_flavour}
		\begin{tabular}{|c|c|c|c|c|}
			\hline
			\multicolumn{2}{|c|}{Observables} & no. of measurements	& scale(s) &  References \\
			\hline
            \multicolumn{2}{|c|}{\bf{Electroweak Precision Observables (EWPO)}}& 13 & $M_Z$ & \cite{ALEPH:2005ab} \\
			\hline
			\multicolumn{2}{|c|}{\bf{PVE and lepton scattering}} & 163 & &  \\ 
			\cline{1-3}
			\multirow{4}{*}{PVE} & $Q_W^{\text{Cs}}$ & 1 & $0.1$ GeV$^*$ & \cite{ParticleDataGroup:2016lqr}  \\
			\cline{2-5}
			& $Q_W^{\text{p}}$ & 1 & $0.1$ GeV$^*$ & \cite{Qweak:2013zxf}  \\ 
            \cline{2-5}
			& $A_{1,2}^{\text{PVDIS}}$ & 2 & $1$ GeV$^*$& \cite{PVDIS:2014cmd} \\
            \cline{2-5}
			& SAMPLE & 1 & $0.2$ GeV$^*$ & \cite{Beise:2004py} \\
			\hline
			\multirow{6}{*}{lepton scattering} & $\nu_\mu\nu_\mu e e$ & 2 & $0.1$ GeV$^*$ & \cite{ParticleDataGroup:2016lqr}  \\
            \cline{2-5}
			& $P_\tau,\,A_P$ & 2 & $58$ GeV & \cite{VENUS:1997cjg}\\
			\cline{2-5}
			& $g_{AV}^{ee}$ in $e^-e^-\to e^-e^-$ & 1 & $0.16$ GeV$^*$ & \cite{ParticleDataGroup:2016lqr}  \\ 
            \cline{2-5}
			&  $A_{\text{FB}}^{\mu,\tau}$ in $e^+e^-\to l^+ l^-$ & 24 & $\sqrt{s}_{\text{LEP}}$ & \cite{ ALEPH:2013dgf, Electroweak:2003ram}\\
            \cline{2-5}
			& $\sigma_{\mu,\tau}$ in $e^+e^-\to l^+ l^-$ & 24 & $\sqrt{s}_{\text{LEP}}$ & \cite{ALEPH:2013dgf, Electroweak:2003ram}\\
            \cline{2-5}
			& $\frac{d\sigma(ee)}{d\cos{\theta}}$ in $e^+e^-\to l^+ l^-$ & 105 & $\sqrt{s}_{\text{LEP}}$ & \cite{ALEPH:2013dgf, Electroweak:2003ram}\\
   			\hline
			\multicolumn{2}{|c|}{\bf{Flavour}} & 37 &  &  \\ 
			\cline{1-3}
			  \multicolumn{2}{|c|}{Differential  $\text{BR}(B\to K\mu\mu)$ (from $14$ GeV)}  & 3  &  & \cite{LHCb:2014cxe}  \\
            \multicolumn{2}{|c|}{Differential  $\text{BR}(B\to K^* \mu\mu)$ (from $14$ GeV)}  & 3  & $m_b$ & \cite{CMS:2015bcy}  \\
            \multicolumn{2}{|c|}{Differential  $\text{BR}(\Lambda_b\to \Lambda \mu\mu)$ (from $15$ GeV)}  & 1  & & \cite{LHCb:2015tgy}  \\
	\cline{1-5}
			\multicolumn{2}{|c|}{$\text{BR}(B\to X_s\mu\mu,\,\mu\mu,\,X_s\gamma,\,K^*\gamma,\,K^{(*)}\Bar{\nu}\nu)$} & 5 & $m_b$ & \cite{BaBar:2013qry,Greljo:2022jac,Misiak:2017bgg,HFLAV:2014fzu,BaBar:2013npw}\\ 
   \multicolumn{2}{|c|}{$\text{BR}(B_s\to \mu\mu,\,\phi\gamma$)} & 2 & $m_b$ & \cite{Greljo:2022jac,Belle:2014sac}
    \\ 
    \multicolumn{2}{|c|}{$\text{BR}(K^+\to\mu^+\nu_\mu,)$}
    & 1 & $m_c^*$ & \cite{ParticleDataGroup:2022pth} 
    \\ 			
    \cline{1-5}
   \multicolumn{2}{|c|}{ $R_K$ and $R_K^*$}
    & 4 & $m_b^*$ & \cite{LHCb:2022vje} 
   \\ 			
   \cline{1-5}
\multicolumn{2}{|c|}{B meson mixing}
    & 2 & $m_b$ & \cite{HFLAV:2016hnz} \\ 
   
    \multicolumn{2}{|c|}{ K meson mixing}
    & 1 & $m_c^*$ & \cite{ParticleDataGroup:2018ovx} \\
    \multicolumn{2}{|c|}{ D meson mixing}
    & 8 & $m_c^*$ & \cite{HFLAV:2016hnz} \\
	\cline{1-5}
		\multicolumn{2}{|c|}{Angular observables in $B\to K^*\mu\mu$ and $\Lambda_b\to\Lambda\mu\mu$ (from $15$ GeV)}  & 16  & $m_b$  & \cite{LHCb:2020lmf,LHCb:2015tgy} \\
    \cline{1-5}
    
    \multicolumn{2}{|c|}{ $\Delta_{\text{CKM}}$}  & 1  & $M_Z$  &   \cite{Gonzalez-Alonso:2016etj,Falkowski:2020pma}  \\
     \hline
		\end{tabular}
	\end{adjustbox}
\end{table}

\FloatBarrier
%----------------------------------------------
\section{Numerical results}
\label{app:num_res}
%----------------------------------------------
We present the correlation matrix of our global fit based on RGE improved LO SMEFT predictions in Figure~\ref{fig:corr_mat}. 
Numerical values for the results of the global fits based on LO, RGE improved LO and RGE improved NLO SMEFT predictions can be found in Table~\ref{tab:fits_numerical}.

\begin{figure}[H]
    \centering
    \includegraphics[scale=0.65]{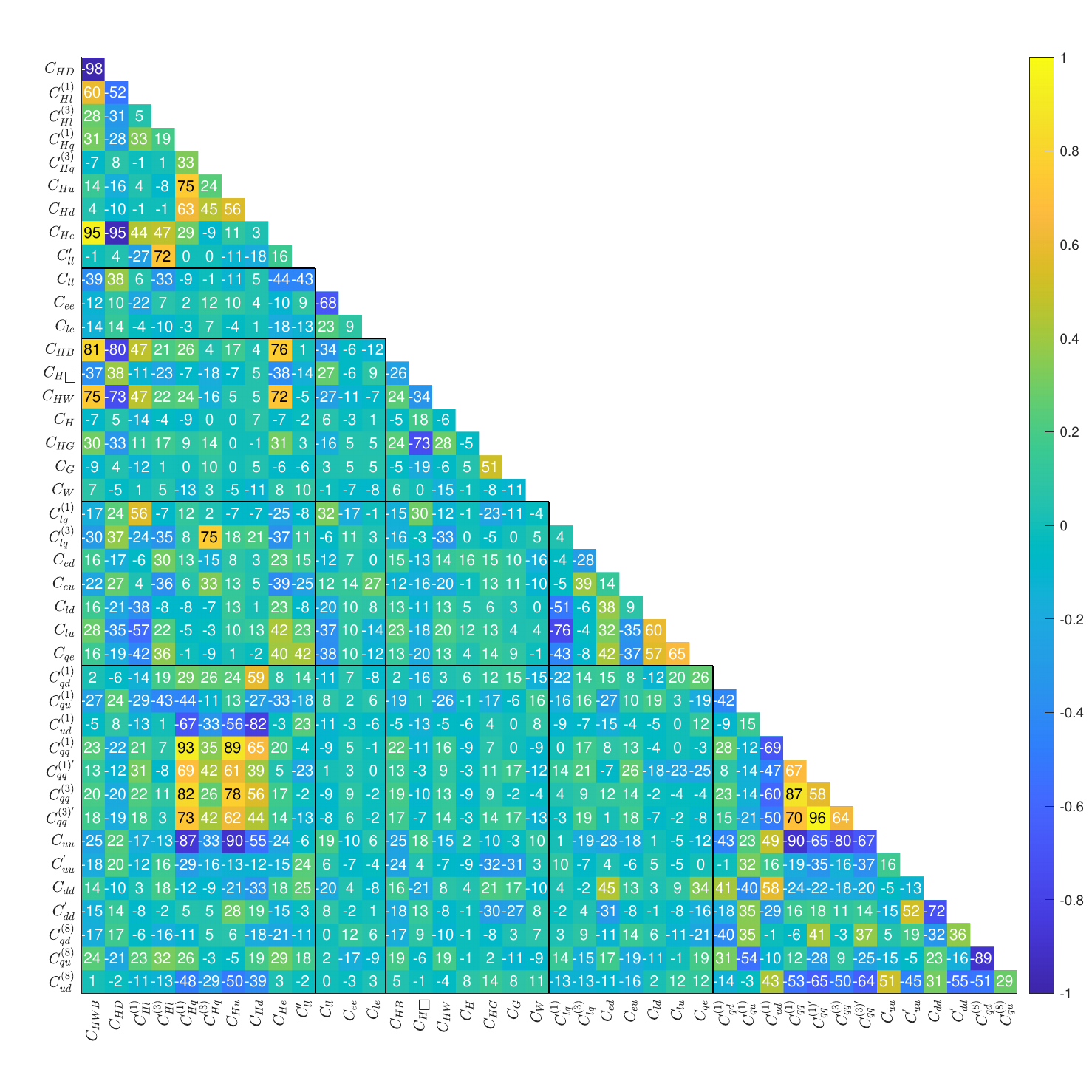}
    \caption{Correlation matrix of the LO global analysis including RGE effects. The numbers in the matrix correspond to the correlations in percent. }
    \label{fig:corr_mat}
\end{figure}

\begin{table}[h]
    \normalsize
    \centering
    \vspace{-0.9cm}
    \begin{tabular}{lccc}
    \toprule
    & \multicolumn{3}{c}{global 95\%CL limits}\\
        coefficient  &  LO  & LO+RGE & NLO+RGE  \\
\hline 
$C_{HWB}$ & $[ -0.14  , \,  0.09 ]$ & $[ -0.16  , \,  0.13 ]$ & $[ -0.14  , \,  0.14 ]$ \\
$C_{HD}$ & $[ -0.24  , \,  0.31 ]$ & $[ -0.39  , \,  0.35 ]$ & $[ -0.4  , \,  0.34 ]$ \\
$C_{Hl}^{(1)}$ & $[ -0.06  , \,  0.069 ]$ & $[ -0.16  , \,  0.04 ]$ & $[ -0.14  , \,  0.05 ]$ \\
$C_{Hl}^{(3)}$ & $[ -0.058  , \,  0.03 ]$ & $[ -0.069  , \,  0.05 ]$ & $[ -0.072  , \,  0.031 ]$ \\
$C_{Hq}^{(1)}$ & $[ -0.11  , \,  0.04 ]$ & $[ -0.92  , \,  0.64 ]$ & $[ -0.8  , \,  0.34 ]$ \\
$C_{Hq}^{(3)}$ & $[ -0.044  , \,  0.042 ]$ & $[ -0.045  , \,  0.054 ]$ & $[ -0.052  , \,  0.043 ]$ \\
$C_{Hu}$ & $[ -0.26  , \,  0.1 ]$ & $[ -0.82  , \,  0.81 ]$ & $[ -0.71  , \,  0.58 ]$ \\
$C_{Hd}$ & $[ -0.81  , \,  0.07 ]$ & $[ -2.7  , \,  2.5 ]$ & $[ -2.4  , \,  0.1 ]$ \\
$C_{He}$ & $[ -0.16  , \,  0.11 ]$ & $[ -0.18  , \,  0.18 ]$ & $[ -0.17  , \,  0.18 ]$ \\
$C_{ll}^\prime$ & $[ -0.072  , \,  0.013 ]$ & $[ -0.074  , \,  0.026 ]$ & $[ -0.078  , \,  0.021 ]$ \\
\hline
$C_{ll}$ & $[ 0.02  , \,  0.23 ]$ & $[ -0.0  , \,  0.22 ]$ & $[ 0.0  , \,  0.21 ]$ \\
$C_{ee}$ & $[ -0.16  , \,  0.01 ]$ & $[ -0.15  , \,  0.03 ]$ & $[ -0.14  , \,  0.03 ]$ \\
$C_{le}$ & $[ -0.025  , \,  0.039 ]$ & $[ -0.023  , \,  0.044 ]$ & $[ -0.025  , \,  0.047 ]$ \\
\hline
$C_{HB}$ & $[ -0.036  , \,  0.042 ]$ & $[ -0.042  , \,  0.064 ]$ & $[ -0.033  , \,  0.07 ]$ \\
$C_{H \square}$ & $[ -1.3  , \,  0.3 ]$ & $[ -1.7  , \,  0.2 ]$ & $[ -1.7  , \,  0.5 ]$ \\
$C_{HW}$ & $[ -0.19  , \,  0.09 ]$ & $[ -0.22  , \,  0.08 ]$ & $[ -0.22  , \,  0.07 ]$ \\
$C_H$ & $[ -10  , \,  5 ]$ & $[ -18  , \,  9 ]$ & $[ -20  , \,  7 ]$ \\
$C_{HG}$ & $[ -0.005  , \,  0.003 ]$ & $[ -0.002  , \,  0.003 ]$ & $[ -0.003  , \,  0.003 ]$ \\
$C_G$ & $[ -0.17  , \,  0.41 ]$ & $[ -0.16  , \,  0.7 ]$ & $[ -0.12  , \,  0.71 ]$ \\
$C_W$ & $[ -0.14  , \,  0.4 ]$ & $[ -0.19  , \,  0.48 ]$ & $[ -0.14  , \,  0.46 ]$ \\
\hline
$C_{lq}^{(1)}$ & $[ -0.64  , \,  -0.01 ]$ & $[ -0.47  , \,  0.02 ]$ & $[ -0.54  , \,  0.07 ]$ \\
$C_{lq}^{(3)}$ & $[ -0.061  , \,  0.053 ]$ & $[ -0.062  , \,  0.068 ]$ & $[ -0.064  , \,  0.064 ]$ \\
$C_{ed}$ & $[ -0.03  , \,  0.88 ]$ & $[ -0.0  , \,  1.3 ]$ & $[ -0.0  , \,  1.2 ]$ \\
$C_{eu}$ & $[ -0.27  , \,  0.33 ]$ & $[ -0.12  , \,  0.33 ]$ & $[ -0.12  , \,  0.37 ]$ \\
$C_{ld}$ & $[ -0.1  , \,  1.8 ]$ & $[ -0.2  , \,  1.4 ]$ & $[ -0.2  , \,  1.6 ]$ \\
$C_{lu}$ & $[ -0.0  , \,  1.2 ]$ & $[ -0.07  , \,  0.97 ]$ & $[ -0.2  , \,  1.1 ]$ \\
$C_{qe}$ & $[ -0.09  , \,  0.88 ]$ & $[ -0.12  , \,  0.59 ]$ & $[ -0.18  , \,  0.57 ]$ \\
\hline
$C_{qd}^{(1)}$ & $[ -730  , \,  670 ]$ & $[ -13  , \,  14 ]$ & $[ -8.3  , \,  4.4 ]$ \\
$C_{qu}^{(1)}$ & $[ -140  , \,  130 ]$ & $[ -3.2  , \,  3.7 ]$ & $[ -0.3  , \,  1.9 ]$ \\
$C_{ud}^{(1)}$ & $[ -170  , \,  220 ]$ & $[ -21  , \,  14 ]$ & $[ -6  , \,  14 ]$ \\
$C_{qq}^{(1)}$ & $[ -6.2  , \,  5.7 ]$ & $[ -3.4  , \,  2.8 ]$ & $[ -3.5  , \,  2.0 ]$ \\
$C_{qq}^{(1)\prime}$ & $[ -0.09  , \,  0.36 ]$ & $[ -0.17  , \,  0.51 ]$ & $[ -0.24  , \,  0.45 ]$ \\
$C_{qq}^{(3)}$ & $[ -0.0  , \,  0.11 ]$ & $[ -0.06  , \,  0.14 ]$ & $[ -0.07  , \,  0.13 ]$ \\
$C_{qq}^{(3)\prime}$ & $[ -0.05  , \,  0.34 ]$ & $[ -0.07  , \,  0.6 ]$ & $[ -0.13  , \,  0.52 ]$ \\
$C_{uu}$ & $[ -6.1  , \,  7.7 ]$ & $[ -3.9  , \,  3.9 ]$ & $[ -3.3  , \,  4.3 ]$ \\
$C_{uu}^{\prime}$ & $[ -1.8  , \,  0.9 ]$ & $[ -1.2  , \,  0.8 ]$ & $[ -1.6  , \,  0.5 ]$ \\
$C_{dd}$ & $[ -160  , \,  130 ]$ & $[ -47  , \,  50 ]$ & $[ -28  , \,  65 ]$ \\
$C_{dd}^{\prime}$ & $[ -63  , \,  88 ]$ & $[ -26  , \,  32 ]$ & $[ -36  , \,  18 ]$ \\
$C_{qd}^{(8)}$ & $[ -6.3  , \,  6.0 ]$ & $[ -4.5  , \,  7.0 ]$ & $[ -5.0  , \,  5.5 ]$ \\
$C_{qu}^{(8)}$ & $[ -1.5  , \,  1.1 ]$ & $[ -1.3  , \,  0.6 ]$ & $[ -1.2  , \,  0.5 ]$ \\
$C_{ud}^{(8)}$ & $[ -12  , \,  7 ]$ & $[ -14  , \,  3 ]$ & $[ -9.8  , \,  7.4 ]$ \\
\bottomrule
    \end{tabular}
    \caption{Numerical results of the global analyses based on LO, RGE improved LO and RGE improved NLO SMEFT predictions.}
    \label{tab:fits_numerical}
\end{table}

\FloatBarrier
\newpage
%%%%%%%%%%%%%%%%%%%%%%%%%%%%%%%%%%%%%%%%%%%%%%%%%%%%%%%%%%%%%%%%%%%%%%%%%%%%%%%%%%%%%%%%%%%%%%%%%%%%%%%%%%%%%%%%%%%%%%%%%%%%%%%%%%%%%%%%%%%%%%%%%%%%%%%%%%%%%%%%%%%%%%
%%%%%%%%%%%%%%%%%%%%%%%%%%%%%%%%%%%%%%%%%%%%%%%%%%%%%%%%%%%%%%%%%%%%%%%%%%%%%%%%%%%%%%%%%%%%%%%%%%%%%%%%%%%%%%%%%%%%%%%%%%%%%%%%%%%%%%%%%%%%%%%%%%%%%%%%%%%%%%%%%%%%%%

\bibliographystyle{JHEP}
\bibliography{bibliography}

\end{document}